\documentclass[prl,aps,twocolumn,showpacs,preprintnumbers,amsmath,amssymb]{revtex4}
\usepackage[dvips]{graphicx}
\usepackage[]{caption}
\usepackage{amsmath}
\usepackage{amssymb}

\usepackage{graphicx}% Include figure files
\usepackage{dcolumn}% Align table columns on decimal point
\usepackage{bm}% bold math
\def\be{\begin{equation}}
\def\ee{\end{equation}}
\def\bea{\begin{eqnarray}}
\def\eea{\end{eqnarray}}

\begin{document}

\preprint{}

\date{\today}

\title{A Study of Structure Formation and Reheating in the D3/D7 Brane 
Inflation Model}

\author{Robert H. Brandenberger$^{1}$} 
\email[email: ]{rhb@hep.physics.mcgill.ca}
\author{Keshav Dasgupta$^{1}$} \email[email: ]{keshav@hep.physics.mcgill.ca}
\author{Anne-Christine Davis$^{2}$} \email[email: ]{A.C.Davis@damtp.cam.ac.uk}

\affiliation{1) Department of Physics, McGill University, Montr\'eal, 
QC, H3A 2T8, Canada}
\affiliation{2) DAMTP, Centre for Math. Sciences, University of
Cambridge, Wilberforce Road, Cambridge, CB3 0WA, U.K.}

\pacs{98.80.Cq}

\begin{abstract}

We study the spectrum of cosmological fluctuations  
in the D3/D7 brane inflationary universe with particular
attention to the parametric excitation of entropy modes
during the reheating stage. The same tachyonic instability
which renders reheating in this model very rapid leads to
an exponential growth of entropy fluctuations during the
preheating stage which in turn may induce a large contribution
to the large-scale curvature fluctuations. We take into
account the effects of long wavelength quantum fluctuations in 
the matter fields. As part of this work, we perform an
analytical analysis of the reheating process. We find that the initial
stage of preheating proceeds by the tachyonic instability channel.
An upper bound on the time it takes for the energy initially stored
in the inflaton field to convert into fluctuations is obtained by
neglecting the local fluctuations produced during the period
of tachyonic decay and analyzing the decay of the residual
homogeneous field oscillations, which  proceeds by 
parametric resonance. We show that in spite of the
fact that the resonance is of narrow-band type, it is sufficiently
efficient to rapidly convert most of the energy of the background fields 
into matter fluctuations.

\end{abstract}

\maketitle

\newcommand{\eq}[2]{\begin{equation}\label{#1}{#2}\end{equation}}

\section{Introduction}

In recent years a lot of interest has focused on the interface area between 
superstring theory and cosmology. The reasons are twofold. Firstly,
cosmology can provide a possible arena to test string theory observationally. 
Secondly, superstring theory may be able to resolve the conceptual problems 
from which the current realizations of scalar field-driven inflation suffer 
(see e.g. \cite{RHBrev} for a discussion). 
In particular, a lot of attention has recently been devoted to
attempts to obtain periods of inflationary expansion of space in
compactifications of superstring theories to four space-time dimensions.
Since such compactifications contain a large number of scalar fields
and since some of those remain massless before supersymmetry breaking,
string theory gives rise to promising candidates for the inflaton,
the scalar field driving the inflationary expansion of space (see
e.g. \cite{Linderev,Cliffrev,JCrev,EvaLiam} for recent reviews of attempts to
obtain inflation from string theory compactifications). 

D-branes have played a particularly important role in the recent
approaches to obtaining inflation from string theory 
\cite{Dvali,Stephon,Shafi,Cliff,JuanGB,KKLMMT}. A particularly
promising model is the D3/D7 brane inflation model, in which the
inflaton is the separation between the two branes in the directions
transverse to both branes \cite{DHHK}. This model features a shift
symmetry \cite{shift} for the inflaton field which ensures that at the 
classical level, the potential is flat along the inflaton 
direction (this shift symmetry is broken under quantum corrections 
\cite{kors}).  
Inflation is induced by supersymmetry breaking terms. Various aspects of this 
model have been studied previously \cite{herd,D37strings,previous}. 

The large number of light moduli fields in string inflation models 
\footnote{These moduli in general could be the complex or K\"ahler 
structure moduli of the internal space. They could 
also be the moduli of the branes in our theory. Once the K\"ahler and 
complex structure moduli are fixed by fluxes \cite{GVW,DRS,GKP}, they could 
be the remaining brane moduli. In this paper, since 
we are not fixing any of the moduli, the light fields could come from all 
the above moduli.}, however,
leads to a new danger: light fields which are not the inflaton are potential
entropy modes and can give rise to entropy fluctuations. The ``curvaton''
\cite{Mollerach,SY,LM,Moroi,LW,Sloth} and ``modulated reheating''
\cite{DGZ,Lev,MR,Uzan,Vernizzi} 
scenarios are examples of this effect. Non-parametric generation of
fluctuations from a global symmetry breaking (also involving an
key isocurvature field) has been investigated in \cite{KRV,Matsuda}.
 
As pointed out in \cite{BaVi,FB2}, these entropy modes
may undergo parametric instability during the initial stages of reheating.
In particular, this instability occurs on super-Hubble (but sub-horizon)
scales. This is possible \cite{FB1}since the background fields carry the 
causal information to scales larger than the Hubble radius. The entropy
fluctuations, in turn, will seed a curvature mode (which we will call
``secondary'' mode). There are two field scalar field toy models in which this
secondary curvature mode dominates over the primary mode, the pure
adiabatic linear perturbation theory mode. In fact \cite{Zibin} the
secondary mode can grow to be larger than unity before back-reaction shuts
off the parametric instability which drives the growth of the entropy
fluctuation. In this case, the model is ruled out by the observed (small)
amplitude of curvature fluctuations on large cosmological scales.

The D3/D7 brane inflation model is a prototypical example in which there
is an entropy mode which can undergo resonance during the initial stages
of reheating, the preheating \cite{JB1,KLS1,JB2,KLS2} phase. In this
paper, we study the growth of this entropy mode and calculate the magnitude
of the induced secondary curvature fluctuations. It turns out that
for reasonable particle physics parameters, the secondary fluctuations
remain in the linear regime at the end of the reheating phase. 
However, their amplitude can be (in the absence of back-reaction
effects) larger than the amplitude of the primary
mode. This would necessitate changing the parameters of the 
model in order to
reproduce the observed magnitude of the large-scale cosmological
perturbations. 

Note that a similar conclusion was recently found 
\cite{Larissa} in another model of brane inflation, the KKLMMT model
\cite{KKLMMT}. In that model, it had already been observed that
due to their large phase space enhancement, second order fluctuations
may give a dominant contribution \cite{BC2}. We should
emphasize that the effect we are calculating in this paper is an effect
which arises in linear cosmological perturbation theory. Both
the linear effects described here and the quadratic processes
discussed in \cite{BC2} and elsewhere (e.g. \cite{Losic,Patrick})
can be operational at the same time. In fact, the enhanced growth
of the linear perturbations which is the focus of this work will
increase the strength of the quadratic processes. We expect
that effects similar to those discussed in \cite{BC2} will also
arise in the D3/D7 system.  

The structure of this paper is as follows: we first revisit the calculation 
of the amplitude of the spectrum of cosmological perturbations in the
D3/D7 inflationary model based on the magnitude of the primary adiabatic
mode alone. Next, we give an approximate but analytical study of reheating 
in the D3/D7 brane inflation scenario. The initial stages of reheating
are governed by a tachyonic decay channel. The same tachyonic resonance
channel leads to exponential growth of super-Hubble entropy modes, which
then in turn induce a secondary curvature mode. The last major section of
this paper focuses on the determination of the amplitude of this mode.

Our analysis of the reheating phase provides results
which are interesting in their own right. 
A new aspects of our analysis is that we allow for an initial
offset in the value of the ``waterfall" field. Such an offset
could arise as a consequence of the back-reaction of long
wavelength \footnote{Long means longer than the Hubble radius.}
fluctuations of the matter fields. Such an offset will suppress
the formation of domains on small scales. If the
offset were sufficiently large, it could prevent the onset of the
tachyonic instability. However, if we take the offset to be
generated by the abovementioned fluctuations, we find
that for most of the realistic space of parameters of the D3/D7 
inflationary model, the resulting values of the field 
lie in the configuration space region for which 
the tachyonic resonance channel \cite{tachyonic} is effective. Thus, the
initial stages of the transfer of the energy in the inflaton field
to matter fluctuations will proceed via the tachyonic instability
in a time interval which is less than the characteristic time it takes
for the matter fields to oscillate about their minima. A substantial
part of the initial inflaton energy is converted to localized
nonlinear fluctuations during the period of tachyonic resonance. 
Neglecting the back-reaction of these fluctuations on the
homogeneous background, we see that a
fraction of order unity of the initial inflaton energy density still
remains in the background when the tachyonic instability shuts off. 
We are interested in demonstrating analytically that the
transfer of energy from the inflaton to fluctuations is effective in
rapidly draining almost all of the energy from the homogeneous fields.
The dominant energy conversion processes are those related to the
interactions of the nonlinear fluctuations. These have been studied
numerically in great detail in \cite{tachyonic}. Since our aim is
to demonstrate analytically that within less than a Hubble time the
energy can be effectively drained from the homogeneous background,
we will neglect the nonlinear interactions and focus on the
evolution of the homogeneous component of the fields once the
tachyonic resonance shuts off (due to our approximation scheme,
we will thus only obtain an upper bound for the decay time). We show that
the decay of the residual homogeneous component of the inflaton field
occurs via a parametric resonance instability 
\cite{JB1,KLS1,JB2,KLS2}. In spite of the fact that the resonance
turns out to be of narrow-band type, it leads to an efficient conversion
of most of the energy from the background inflaton field to matter 
fluctuations.

One of the advantages of the D3/D7 brane inflation model is that the
matter fields are directly associated with the branes whose separation
provides the inflaton field\footnote{Although not directly related to the main theme of the 
paper, we would like to point out that some recent papers \cite{beasly} have shown how to 
get standard model with chiral fermions in a D3/D7 set-up from F-theory. Of course one need to 
change our background manifold to a certain Del-Pezzo surface to accomodate the standard model. 
It would be interesting to realise D3/D7 inflationary dynamics with this manifold.}. 
As a consequence, there is a direct coupling
between the inflaton field and the matter fields, rendering it easy, as
we will see, to obtain efficient reheating. This is a feature the
D3/D7 brane inflation model shares with the D4/D6 model in which
reheating was analyzed in \cite{Easson} and with the brane-antibrane
inflation models in which reheating was studied in \cite{CFM,BC} (see
also \cite{TM}), but contrasts with the
situation in inflationary models in which the inflaton dynamics and
the standard model fields live in different throats, in which case
efficient reheating is more difficult to obtain (see e.g. 
\cite{BBC,KY,DSU,FMM,CT,PL}).

The outline of this paper is as follows. In the following section, we review 
the D3/D7 brane inflationary model, with particular emphasis on the form of 
the potential for the inflaton field. In Section 3 we discuss the
vacuum manifold of the model, and in Section 4 we revisit the
calculation of the spectrum 
of cosmological perturbations. Section 5 focuses on the dynamics of the
inflaton and of the other relevant scalar fields of the model before the
onset of reheating, introduces the conditions for effectiveness of
tachyonic resonance and demonstrates that for most of the relevant parameter
space of the theory, the tachyonic resonance condition is satisfied.
However, the tachyonic instability turns off before most of the energy
of the inflaton field is released as matter fluctuations. To obtain
an upper bound on the time it takes for the inflaton energy stored in
the residual oscillations to further dissipate, we study the later
dynamics of the homogeneous field components which
proceeds via the less efficient parametric
resonance channel (Section 6). Section 7 is devoted to the computation
of the secondary curvature fluctuation mode, the mode induced by the
entropy fluctuations.  

\section{The D3/D7 Brane Inflation Model}

The simplest way \cite{DHHK} 
to construct a D3/D7 inflationary model in type IIB theory 
is to place a D3 brane (which we take to be in the $0 - 3$ directions)
parallel to a D7 brane (which is in the $0 - 3,6 - 9$ directions)
in flat space-time and break the resulting 
${\cal N} =2$ supersymmetry completely by switching on non-primitive gauge
fluxes on the D7 brane. The BPS condition is broken, and the D3 brane starts 
to move towards the D7 brane. The distance between the D3 brane and the 
D7 brane in the $4,5$ plane appears on the 
world-volumes of the branes as a complex scalar $S$, 
whose value changes as the distance between 
the branes changes. This scalar field is identified with the inflaton, and the 
motion of the D3 brane towards the D7 brane is the {\it slow-roll} of hybrid 
inflation. 

Once the D3 brane comes close to the D7 brane, the open string between the 
branes become tachyonic because of the non-primitive fluxes on the seven 
brane. This is the top of the hybrid inflationary scenario where one 
would expect a tachyonic phase transition. 
The D3 brane then falls onto the D7 brane and dissolves as a non-commutative 
instanton \cite{DHHK}. The non-commutative instanton is {\it not} 
point-like and therefore the final configuration
of a dissolved instanton is non-singular \cite{instanton}. 
One can also argue that in the end 
${\cal N} = 2$ supersymmetry can be restored.  

This simple picture has many inherent problems. Firstly, being non-compact, 
the gravitational solution is ten dimensional and a direct dimensional 
reduction will give us vanishing Newton's
constant in four dimensions. One way out is to compactify the other six 
dimensions. Secondly, once we compactify, new problems have to be
faced. An immediate issue that on a compact space, charge neutrality must 
be enforced, which requires introducing additional elements to the
construction. Next, the problem of moduli stabilization arises,
because without stabilizing the moduli,
the system will decompactify back to the ten dimensional solution via the 
so-called Dine-Seiberg runaway \cite{DiS}.
Thirdly, one needs to avoid an over-production of cosmic strings when the D3 
dissolves in the D7 brane. 
 
Ways to solve these problems were addressed in the papers 
\cite{DHHK, previous} where the proposal was to 
consider a warped solution of the form 
\be \label{warped}
ds^2 = e^{A(y,t)} ds^2_{0123} + e^{B(y,t)} g_{mn} dy^m dy^n \, ,
\ee
where $A,B$ are the warp factors and $y^m$ are the coordinates of the 
internal space. In \cite{DHHK} the 
metric $g_{mn}$ was chosen to be the metric of $K3 \times T^2/Z_2$ with 
$Z_2$ being a combination of
an orbifold and an orientifold operation (see \cite{DHHK} for details). 
Both the D3 and the D7 are 
points on the ${\bf P}^1 \equiv T^2/Z_2$, and the motion of the D3 is 
trigerred by three form fluxes 
$H_{NS}$ and $H_{RR}$\footnote{Observe that once $B_{NS}$ is dualised to form gauge field on the world volume of 
D7 brane, this would trigger motion of the D3 brane towards D7 brane.}. 
The D3 brane is oriented along $x^{0,1,2,3}$ and 
therefore is also a point on the 
K3 manifold. The D7 brane wraps the K3 and is oriented parallel to the D3 
as we discussed before. Supersymmetry is broken either by making 
\be \label{susy}
iG \ne \star G \, ,
\ee
where $G = H_{RR} + \tau H_{NS}$, or by switching on ISD 
(1,2) or (0,3) forms (recall that ISD (1,2) fluxes also break susy. For more detailed analysis of this, see the recent 
paper \cite{DFKS}).  
The Hodge star $\star$ is on the internal space and 
\eqref{susy} gives rise to non zero supersymmetry breaking potential
because of its IASD property. 

Once supersymmetry is broken, there would be non-trivial potential between the 
branes. The complete picture can be worked out easily from the low energy 
type IIB  
supergravity lagrangian in the presence of D-brane sources in the following 
way (see also \cite{D37strings}):
\bea \label{comlag}
&& S\,= \, \frac{1}{\kappa_0^2}\int d^{10}x \left[\sqrt{-G_{10}} ~R ~+ ~ 
\frac{G\wedge \ast{\bar G}}{{\rm Im}~\tau} + ..... \right]
+ \nonumber \\ 
&& + T_3 \int d^4 x \sqrt{-G_4} \Biggl[\frac{1}{4} (F_W)^2  + \frac{1}{4} (F_W')^2 + |\partial S|^2 + 
|\partial S'|^2 + \nonumber \\
&& + \Bigg(c_0 \int_{\rm K3} {\cal F}_- \wedge \ast {\cal F}_-\Bigg) 
+ |D_\mu \chi|^2 + 2 g^2 |S|^2 |\chi|^2 + \nonumber \\
&& + \frac{g^2}{2} (\chi^\dagger \sigma^A \chi)^2 - \Delta V\Biggr] + {\rm non~perturbative ~terms}
\eea  
where the dotted terms involve axio-dilaton plus other higher form 
interactions, $c_0$ is a constant and 
${\cal F}_- \equiv {\cal F} - \ast {\cal F}$ with ${\cal F} = F_W'-B$. 
where $F_W$ is the gauge field on the D3 brane, $F_W'$ is the gauge field on the D7 brane, and ($S, S'$) are the 
corresponding scalars on these branes. For more details see 
\cite{D37strings}.   

The first term in eqn. \eqref{comlag} when reduced to four dimensions give 
rise to the kinetic terms for the 
complex and K\"ahler structure moduli. The second term is the potential term 
for the moduli:
\be \label{secterm}
\frac{1}{\kappa_0^2}\int d^6 x \frac{G\wedge \ast{\bar G}}{{\rm Im}~\tau} \equiv e^{K}\left(|DW|^2 - 3|W^2|\right)
\ee
and therefore fixes (at least) all the complex structure moduli 
and the axio-dilaton (the $3|W^2|$ term goes away for the no-scale models).

The second line in eqn. \eqref{comlag}
is related to the kinetic terms for the gauge fields on the D-branes and the 
inflaton $S$ (if we ignore the scalar $S'$). The other scalar doublet $\chi$ 
denote the hypermultiplet scalars with the
corresponding D-term and F-term potentials being given in the third and the 
fourth lines of eqn. \eqref{comlag}. The 
potential $\Delta V$, whose
value will be specified later, denotes the additional attractive potential 
between the D3 and the D7 brane. Finally, the 
non-perturbative potentials stabilize the remaining K\"ahler structure moduli. 

The above picture is {\it almost} complete except for an important subtlety. 
The potential that 
fixes the complex structure moduli in eqn. 
\eqref{secterm} also gives a large mass to the inflaton field $S$ 
because the K\"ahler potential $K$ in eqn. 
\eqref{secterm} picks up a dependence on the inflaton \cite{witten85,KKLMMT}. 
The potential 
for the inflaton then becomes very steep and consequently do not allow slow 
roll to happen in generic models. 
In the D3/D7 model this could be avoided (at least at the tree level) by 
allowing a shift symmetry, but this symmetry 
is broken by the $\Delta V$ term in \eqref{comlag}. Thus a generic model of 
inflation from string theory is  
very complicated (see \cite{BCDF,mcall,krause} for some recent developments) 
and since our aim here is to make the 
first step in understanding structure formation and reheating after inflation, 
we will ignore the subtleties associated
with moduli stabilization. 

Therefore,
from the analysis presented above in eqn. \eqref{comlag} we will only 
concentrate on the F-term and D-term potentials
for the hypermultiplet scalars (which we will henceforth label as $\phi_+$ and 
$\phi_-$ instead of the complex doublet 
$\chi$)
and the inflaton $S$, and assume that the contributions from the 
background fluxes 
given in eqn. \eqref{secterm} and the non-perturbative contributions fix most of the moduli.
Note however that, as we mentioned earlier, there are other contributions to the inflaton potential 
coming from the K\"ahler potential $K$ in \eqref{secterm}. For the simple analysis that we 
present here, we will ignore this contribution. More details on this will appear elsewhere.    

Looking from the D3 point of view, the complex scalar $S$ that parametrises 
the ${\bf P}^1$, and a set of complex scalars $\phi_+$ and $\phi_-$
arising from the hypermultiplet-string between the branes give rise to the 
following potential:
\bea \label{pot1}
&& V(S, \phi_+, \phi_-) \, = 
\, 2 g^2 |S|^2 \bigl(|\phi_+|^2 + |\phi_-|^2 \bigr) \nonumber \\
&&+ 2 g^2 \Bigl| \phi_+ \phi_- - {{\zeta_+} \over 2}\Bigr|^2 
+ {{g^2} \over 2} \bigl( |\phi_+|^2 - |\phi_-|^2 - \zeta_3 \bigr)^2 
\eea
plus the additional $\Delta V$ term that we will add later. Here we have taken 
a more generic case by switching on $\zeta_+$ and $\zeta_3$ as the FI terms to break 
susy. 

There is also another subtlety related to 
the issue of moduli stabilization that we should mention now.
Due to background NS and RR fields, one might expect that 
all the complex structure moduli should be naturally fixed via a GVW potential 
\cite{GVW} (see also \cite{DRS, GKP}). But our background is time-dependent, 
so the issue of fixing moduli 
should be addressed at any {\it given} time. Similarly, one might also expect 
that the supergravity no-scale structure will be
broken by allowing a gaugino condensate along the lines of \cite{KKLT, DDFGK}. 
These problems are not 
yet solved in a time dependent background, and an attempt to address these 
issues along the lines of \cite{previous} is under way (see also
\cite{Brax} for another recent study of 
moduli corrections to D-term inflation).  

Finally, the issue of cosmic strings can be addressed by going to the full 
non-perturbative F-theory 
picture of the D3/D7 system (see \cite{D37strings,lumps} and the second 
paper of \cite{previous}). From F-theory 
it is easy to generate {\it global} symmetries, and an $SU(2)$ global 
symmetry can convert the cosmic strings
to semi-local strings of Achucarro-Vachaspati \cite{AV, UAD}
type (see also \cite{Burrage}). In terms of 
the D3/D7 system, this is simply a
doubling of the number of D7 branes. These semi-local strings reduce the 
severity of the problems generated by cosmic strings \cite{UAD} (they
do not completely eliminate the problems, as a recent study
\cite{Jon} has shown).

\section{The Vacuum Manifold of the Model}

From the previous section, we see that the potential which describes
the dynamics of the complex inflaton field $S$ (the separation of the two
branes in the 4-5 plane) and of the two complex scalar fields 
$\phi_+$ and $\phi_-$ associated with the lowest energy modes of the 
strings stretching between the branes is given by (\ref{pot1}).

The vacuum manifold ${\cal M}$ is the set of field configurations which
minimize the potential. The potential has been normalized such that
the minimal value of the potential is $V = 0$. To be at the minimum of
the potential, the inflaton field must have relaxed to $S = 0$. A
special case which we will focus most of our attention on is the
case $\zeta_+ = 0$ with $\zeta_3 \neq 0$. Introducing radial and angular
coordinates with each $\phi$ field:
\bea
\phi_+ \, &=& \, \rho_+ e^{i\varphi_+} \\
\phi_- \, &=& \, \rho_- e^{i\varphi_-}
\eea
we see that the conditions to have $V = 0$ become
\bea
\rho_+ \rho_- \, &=& \, 0 \nonumber \\
\rho_+^2 - \rho_-^2 - \zeta_3 \, &=& \, 0 \, .
\eea
The vacuum manifold ${\cal M}$ is given by
\be
{\cal M}\, : \,\,\,\, \rho_+ = \sqrt{\zeta_3}, \,\,\, \rho_- = 0 \, .
\ee
It is the $S^1$ manifold discussed in \cite{previous}. There are
no non-compact flat directions.

Another interesting special case occurs if $\zeta_3 = 0$ and $\zeta_+ \neq 0$.
In this case, setting the potential to zero yields the conditions
\bea
|\phi_+|^2 - |\phi_-|^2 \, &=& \, 0 \nonumber \\
|\phi_+ \phi_- - {{\zeta_+} \over 2}| \, &=& \, 0 \, .
\eea
In this case, the vacuum manifold is the circle given by
\be
{\cal M} \, : \,\,\,\, \rho_+ = \rho_- = \sqrt{\zeta_+ / 2}, \,\,\,
\varphi_+ = - \varphi_- \, .
\ee
There are no non-compact flat directions.

In the general case there are no flat directions, either, and the vacuum
manifold is given by
\bea
\rho_- \, &=& \, {1 \over {\rho_+}} {{\zeta_+} \over 2} \nonumber \\
\rho_+^2 \, &=& \, {{\zeta_3} \over 2} + 
{1 \over 2} \sqrt{\zeta_3^2 + \zeta_+^2}
\nonumber \\
\varphi_+ &=& - \varphi_- \, .
\eea

In the following, we will focus mostly on the special case $\zeta_+ = 0$.

\section{Spectrum of Cosmological Fluctuations}

In this section we will review the calculation of the amplitude
of the primary adiabatic perturbations. This is the mode which is
usually considered in the literature and which is used to fix
the model parameters. In Section 7 we will turn to a discussion of
the entropy mode.

In the D3/D7 brane inflation scenario, inflation takes place while
the branes are widely separated, i.e. while $|S|$ is large. From (\ref{pot1}) 
it follows immediately that for large values of the inflaton, the scalar
fields $\phi_+$ and $\phi_-$ will vanish at the classical level (their
quantum fluctuations will play an important role in the following section).

Before turning on any supersymmetry breaking, and even after
introducing the supersymmetry-breaking parameters $\zeta_+$ and $\zeta_3$
at the tree level, the configuration of branes
is BPS, there is no force between the branes, and $S$ will remain constant.
One loop supersymmetry breaking terms will result in an additional one-loop 
contribution $\Delta V$ to the potential
\be \label{pot2}
\Delta V \, = \, 
{{g^4} \over {16 \pi^2}} \zeta_3^2 
~{\rm ln}\Biggl({{|S|^2} \over {S_c^2}}\Biggr) \, .
\ee
In the language of supersymmetric effective field theory, this
term corresponds to D-term supersymmetry breaking \cite{Dterm}.
We are omitting the quartic contribution to the symmetry breaking
potential which corresponds to F-term supersymmetry breaking. In the 
language of D3/D7 system, the above potential appears naturally from the 
supergravity solution of a D3-brane separated from a D7-brane by a radial distance $r$ as \cite{DHHK}:
\be \label{sugrapot}
\Delta V \, \propto \, {\rm arctan}^2 ~{\cal F}_- ~{\rm ln}~\Biggl({r^2\over \Lambda^2}\Biggr) \, ,
\ee
where ${\cal F}_-$ is related to the D7 brane gauge fluxes. Therefore the 
identification of $\zeta_3$ with some components of the gauge fluxes on the 
D7 brane is consistent with expectation. We also see that $S_c$ in (\ref{pot2}) is
related to the string scale $\Lambda$ in (\ref{sugrapot}).

For small values of the string coupling constant $g$, this extra
term in the potential will lead to a slow rolling of the inflaton 
field $S$ towards $S = 0$. Initially, the values of $\phi_+$ and
$\phi_-$ will vanish. Note that the presence of (\ref{pot2}) lifts
the flat direction, preferring the value $\phi_- = 0$. For a 
sufficiently small value of $S$, a tachyonic instability of the
field $\phi_+$ will set in. The critical value $|S| = S_t$ below
which this instability appears is given by
\be \label{dis}
S_t \, = \, \sqrt{{{\zeta_3} \over 2}} \, , 
\ee
the condition coming from setting the second derivative of the
potential (\ref{pot1}) to zero at $\phi_+ = 0$. From direct
open string calculation this is given by 
\be \label{dis2}
S_t \, = \, \sqrt{\pi\alpha'{\rm arctan}~{\cal F}_-\over 2} \, .
\ee
Again we see that the 
identification of $\zeta_3$ with some components of the gauge fluxes on the 
D7 brane is consistent with expectation. 

Since there is no instability in the $\phi_-$ direction, $\phi_-$ will
remain zero, again modulo quantum uncertainties which may become
important during the later stages of preheating. We will therefore,
for the moment, set $\phi_- = 0$ and consider the dynamics of
the reduced configuration space consisting of the inflaton and $\phi_+$.

The inflationary scenario realized here is a specific case of
hybrid inflation \cite{hybrid}, and more precisely of
supersymmetric hybrid inflation \cite{Dterm} (see also \cite{Pterm}).
Whereas it is the field $S$ which is slowly
rolling and producing the density fluctuations, it is the potential
energy of the field $\phi_+$ which generates inflation. Fluctuations
in general hybrid inflation models have been studied before
(see, in particular, \cite{Renata,Pterm}).

Using the standard theory of cosmological perturbations in inflationary
cosmology (see e.g. \cite{MFB,LR} for comprehensive review articles
and \cite{RHBrev2} for a pedagogical introduction) we obtain as
amplitude $\delta_H$ of the spectrum of scalar metric fluctuations
\be \label{amp1}
\delta_H \, \simeq \, 
{1 \over {5 \pi \sqrt{3}}}
{{V^{3/2}} \over {m_{pl}^3 \partial V / \partial S}} \, ,
\ee
where $m_{pl}$ is the reduced Planck mass. In terms of the D3/D7 model, this 
could be derived from the fluctuation of the background metric when the D3 brane
moves towards the D7 brane. 
Now,
making use of (\ref{pot1}) and (\ref{pot2}) we obtain (see also \cite{D37strings})
\be \label{amp2}
\delta_H \, \simeq \, {{4 \pi} \over {5 \sqrt{6}}} 
{{\zeta_3 S} \over {g m_{pl}^3}} \, .
\ee
This is to be evaluated at the value of the inflaton $S$ (we will from
now on omit the absolute value sign for this field) when modes which are
being observed today on large scales exit the Hubble radius during the
period of inflation. For high scale inflation this happens $N \simeq 50$
Hubble expansion times before the tachyonic instability develops, at a
value $S = S_N$. Assuming that the acceleration of $S$ during
inflation is negligible, then $S_N$ is given by
\be \label{Svalue}
S_N^2 \, = \, S_t^2 + N g^2 {{m_{pl}^2} \over {4 \pi^2}} \, ,
\ee
where ${\dot S}$ is evaluated at $S = S_t$ (we solve the scalar field equation of motion
working backwards in time assuming constant ${\dot S}$ to obtain \eqref{Svalue}).

Provided that
\be \label{ineq}
{{N g^2} \over {2 \pi^2}} \Biggl({{m_{pl}} \over {\zeta_3^{1/2}}}\Biggr)^2
\, < \, 1 \, ,
\ee
the second term on the right hand side of (\ref{Svalue}) is smaller
than the first. In this case, (\ref{amp2}) yields
\be \label{amp4}
\delta_H \, \simeq \, {{2 \pi} \over {5 \sqrt{3}}} {1 \over g} 
{{\zeta_3^{3/2}} \over {m_{pl}^3}} \, .
\ee
On the other hand, if the inequality in (\ref{ineq}) is reversed,
then the result becomes
\be \label{amp5}
\delta_H \, \simeq \, {2 \over {5 \sqrt{6}}} \sqrt{N} 
{{\zeta_3} \over {m_{pl}^2}} \, .
\ee

\section{Scalar Field Dynamics and Conditions for Tachyonic Resonance}

The cosmological evolution starts with a separation between the branes
which is large on string scale. At large values of $|S|$, the coupling
between $S$ and the string fields $\phi_+$ and $\phi_-$ constrains
the latter to be zero, modulo classical and quantum fluctuations.

Neglecting the classical and quantum fluctuations of $\phi_+$ and $\phi_-$,
the cosmological evolution is as follows: the inflaton field $S$ rolls
down the ``valley" of its potential until a tachyonic instability for
$\phi_+$ sets in. This occurs at the value $|S| = S_t$ given by (\ref{dis}).
%%
%\be
%S_t \, = \, \bigl( {{\zeta_3} \over 2} \bigr)^{1/2} \, .
%\ee
%%

At that point, the evolution will take place in the complex two-dimensional
field space of $S$ and $\phi_+$. There is no instability in $\phi_-$
direction, and hence $\phi_-$ will remain zero. At the background level,
the term proportional to $|\phi_+ \phi_-|$ induces a linear confining
potential for $\phi_-$.

Let us now focus on the potential in $\phi_+$ direction for a fixed value
of $S$. This potential has a negative effective mass (i.e. a
tachyonic instability) for small values of $|\phi_+|$ only. For larger
values, the effective square mass is positive, leading to oscillations
about the minimum of the potential.

The phases of $S$ and $\phi_+$ do not play an important role in most
of our considerations, and hence we will set them to zero in the
following unless indicated otherwise. This will simplify the
notation. We will briefly return to the issue of these phases
in the concluding section.

A new aspect of our work is the inclusion of the possibility that
the $\phi_+$ field has an offset from the symmetric point
$\phi_+ = 0$ at the end of the period of inflation \footnote{Note that such an
offset has been discussed in a different context in \cite{Rachel}.}. Such an
offset in any given Hubble patch will in general be produced by
the fluctuations of the field on super-Hubble scales
\cite{Starob,BR}. The offset will suppress the ability of
small-scale quantum vacuum fluctuations of $\phi_+$ to produce
topological or non-topological defects on small length scales
at the beginning of the resonance process \footnote{A
quantitative study of the issue is currently in progress.}. Such
defect formation and the subsequent non-linear interaction of the
defects was seen in previous work \cite{tachyonic} to be the most
important aspect of reheating in hybrid-type inflation models
like the one we are considering. In our analytical study
of reheating, we will assume that
the effect of the small-scale vacuum fluctuations is sub-dominant
compared to the effect of the offset. Even if the small-scale fluctuations
dominate the reheating dynamics, they will unlikely influence the
evolution of the large-scale metric fluctuations which is the focus
of most of our interest.

The first question to ask concerning the effects of the offset is
whether it can prevent the onset of the ``tachyonic'' (or ``negative
coupling'') \cite{GB2,tachyonic2} resonance. For the sake
of concreteness, we will assume that the offset is generated by
the long wavelength fluctuations of $\phi_+$ which were generated
during the period of slow-roll inflation and check to see if
they are sufficiently small such that, when 
$S$ reaches the onset of the instability, $\phi_+$ will in fact be
in the region of tachyonic evolution. If this is the case then,
as discussed in \cite{tachyonic}, the
energy transfer from the coherently evolving fields to matter
fluctuations (the ``preheating" phase of the reheating process) will
be very rapid and complete before oscillations of $\phi_+$ and
$S$ can set in. If, on the other hand, the fluctuations of
$\phi_+$ are sufficiently large, then there will be no tachyonic
preheating phase, and the energy transfer from the inflaton to
the matter fields will proceed by the parametric instability
channel first discussed in \cite{JB1} (see also \cite{DK}) and 
discussed comprehensively in \cite{KLS2}.
 
Before estimating the magnitude of the fluctuations of $\phi_+$,
we note that the slow rolling of the inflaton field $S$ will
prevent the onset of the instability of $\phi_+$ until
\be \label{endslow}
\Bigl|{{\partial V} \over {\partial \phi_+}}\Bigr| \, \geq \, 
\Bigl|{{\partial V} \over {\partial S}}\Bigr| \, .
\ee
Before this condition is satisfied, the evolution of the dynamical
system in the $S - \phi_+$ plane corresponds to a horizontal
straight line (the solid curve in Figure 1).

\begin{figure}
%\centerline{\epsfxsize=3in\epsfbox{spacetime.eps}}
\includegraphics[height=6cm]{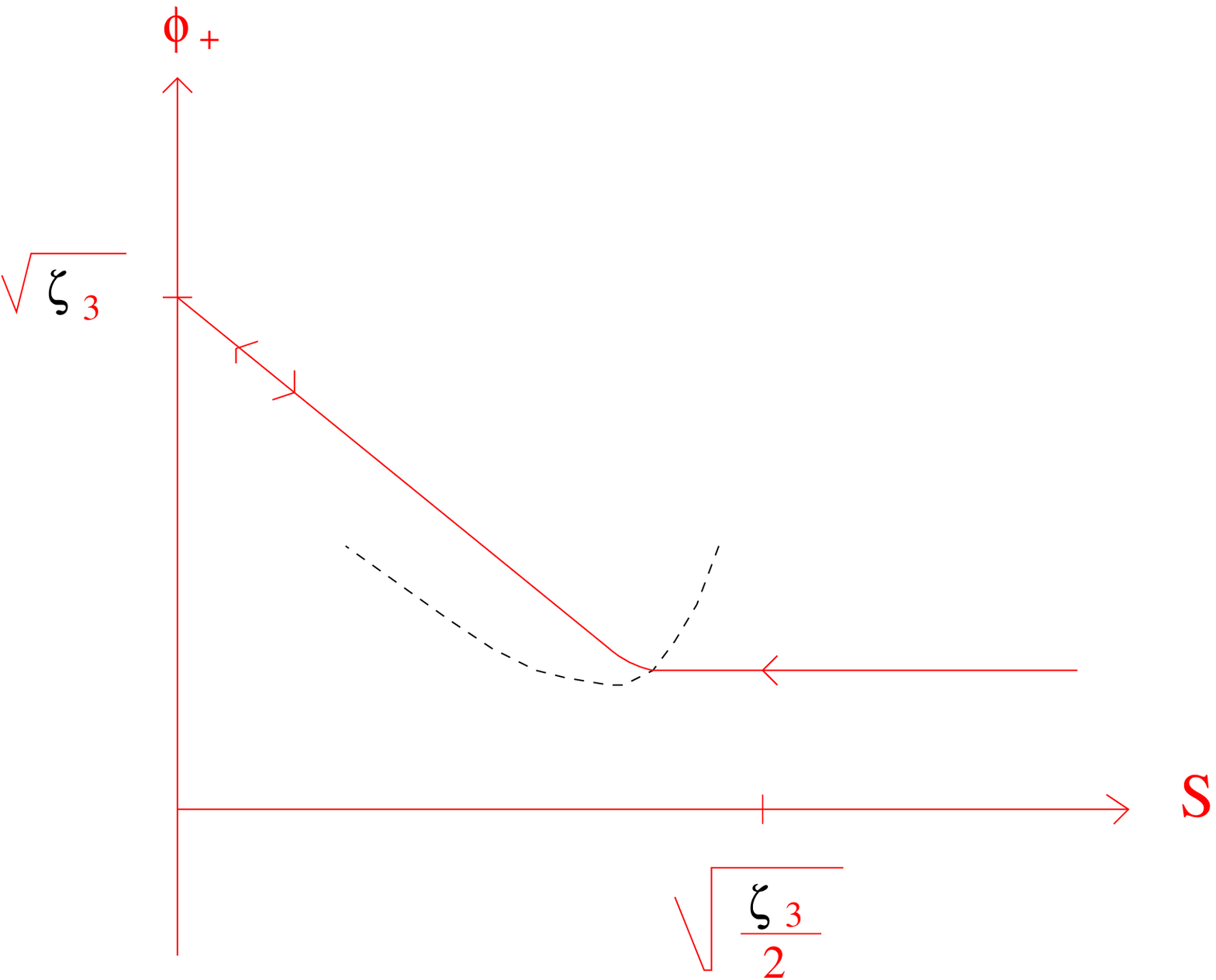}
\begin{caption}
{\small Sketch of the evolution of the fields $S$ and $\phi_+$
in configuration space. The solid line represents the evolution
of the background fields. The cosmological evolution begins in
the region of slow-roll (to the right and below the dashed curve).
The slow-rolling ends when the dashed line is reached, after which
a period of quasi-periodic oscillations about the ground state
$S = 0$ and $\phi_+ = \zeta_3$ sets in.}
\end{caption}
\label{fig:1}
\end{figure}

We next find the region of phase space where the dynamics is
dominated by the instability in the $\phi_+$ direction.
{F}rom the potentials (\ref{pot1}) and (\ref{pot2}) (with $\phi_- = 0$)
it follows that the phase space curve where the two
derivatives in (\ref{endslow}) are equal is given by the following relation:
\be \label{actcond}
\phi_+ \Biggl(\frac{2S^2}{\zeta_3} - 1\Biggr) + \frac{\phi_+^3}{\zeta_3} 
+ \frac{2S}{\zeta_3} \phi_+^2 \, = \,  
- \frac{g^2 \zeta_3}{16\pi^2 S}
\ee
Above this curve, the dynamics is dominated by the $\phi_+$ instability.
In the limit where $\phi_+$ is small, we can simplify the potential such 
that the equation relating $\phi_+$ and $S$ 
has no quadratic or cubic powers of $\phi_+$. If we also ignore the phase 
$\varphi_+$ of $\phi_+$, and work in the limit $S \ll \zeta_3^{1/2}$ 
(this approximation is good except at the beginning of the resonance)
we get the 
following approximate relation:
\be \label{cond3}
\phi_+ ~ \equiv ~\rho_+ \, \simeq \, 
{{g^2} \over {16 \pi^2}} \zeta_3 {1 \over S} \, .
\ee
The minimal value of $\phi_+$ is taken on when $S \sim \zeta_3^{1/2}$,
which yields a value of about 
$\rho_+ = g^2 / (16 \pi^2) \zeta_3^{1/2}$.
The curve where the condition (\ref{cond3}) is satisfied is 
represented by the dashed curve in Figure 1.

The condition for efficiency of tachyonic reheating is
\be \label{tachcond}
{{\partial^2 V} \over {\partial \rho_+^2}} \, \leq \, 0 \, ,
\ee
evaluated at the phase space point where the phase space
trajectory of the background dynamics of our $S - \phi_+$ system
crosses the dashed curve given by (\ref{cond3}). This would mean
\be \label{phiplusrel}
\rho_+ \, \leq \, \sqrt{\frac{\zeta_3}{3} - \frac{2}{3} S^2}
\ee
{}From here it is
clear that the conditions
(\ref{tachcond}) and \eqref{phiplusrel} correspond to
\be \label{cond4}
 \rho_+ \, < \, \sqrt{{\zeta_3} \over 3} \, ,
\ee
which along the trajectory of Fig. 1 is automatically satisfied
for small values of $g$ if $|S| < S_t$.

We must now check whether the condition (\ref{tachcond}) is satisfied
including the effects of fluctuations of $\phi_+$. Thus, we
must calculate the magnitude of the root mean
square fluctuations of $\phi_+$. Since at the beginning of the
cosmological evolution, for large values of $S$, the valley of
the potential is very steep, and since during inflation all
pre-existing classical fluctuations are red-shifted, we expect the
classical fluctuations to be negligible. Hence, we focus on the
quantum fluctuations. The quantum vacuum fluctuations averaged
over a Hubble volume are given by integrating the Fourier space
quantum vacuum fluctuations for all comoving modes which are 
super-Hubble \cite{Starob,BR}:
\be \label{expr}
<\rho_+^2>_Q \, \sim \, \int^{H_c(t)} d^3k ~|\phi_+(k)|_Q^2 \, ,
\ee
where the subscript $Q$ refers to the fact that we are computing
a quantum vacuum expectation value, and the $H_c$
is the comoving Hubble constant. 
While the field $S$ has a value much larger than
$S_t$, the effective mass for the waterfall field $\phi_+$
will be large, and the vacuum fluctuations small. However, if the
effective mass remains smaller than the Hubble rate $H$ for
several Hubble expansion times before the onset of the
tachyonic instability, then the field fluctuations will approach
the values obtained for a massless field. It can easily be
verified that the above condition for ``effective masslessness''
will be satisfied if (\ref{ineq}) holds. 
In this case, one obtains
\be \label{expr2}
<\rho_+^2>_Q \, \sim \, \int^{H_c(t)} d^3k ~|\phi_+(k)|_{Q,m=0}^2 \, ,
\ee
where the subscript now indicates that we are taking the quantum
fluctuations for a massless scalar field. 
Since in an expanding universe
it is the field $a(t) \phi$ which has canonical normalization,
evaluating the right-hand side of (\ref{expr}) gives
\be
<\rho_+^2>_Q \, \sim \, \int^{H_c(t)} \frac{d^3k}{2 a^2 k} \,
\simeq \, H^2(t) \, .
\ee
The above estimate is based on the same ideas which lead to
stochastic inflation \cite{Starob}.

The Hubble expansion rate at the end of the slow-roll period is
\be \label{Hubble}
H \, = \, g {{\zeta_3} \over {\sqrt{3} m_{pl}}} \, .
\ee
Plugging the value $\rho_+ = H$ with the above value of $H$
into (\ref{cond4}), yields the following condition on the
parameters of our model for which the reheating dynamics
will go through a phase of tachyonic instability:
\be \label{cond5}
g {{\zeta_3^{1/2}} \over {m_{pl}}} \, < \, 1 \, .
\ee

We will initially choose the parameters of the D3/D7 brane inflation
model such that the direct adiabatic fluctuations
produced during inflation, whose magnitude was discussed in the
previous section, provide the primordial fluctuations observed
today. It is now straightforward to check, making use of
(\ref{cond5}) and (\ref{amp4}) or (\ref{amp5}), whether tachyonic
resonance preheating will occur.

Let us consider some characteristic parameter values. Scales which are being
observed today leave the Hubble radius about 50 e-foldings before
the end of inflation. Hence, we take $N = 50$. In the range of
values of $g$ for which the inequality (\ref{ineq}) is violated,
it follows from (\ref{amp5}) that for a COBE-normalized \cite{COBE}
value of $\delta_H \simeq 10^{-5}$, we require
\be
{{\zeta_3^{1/2}} \over {m_{pl}}} \, \simeq \, 10^{-2.5} \, ,
\ee
and that hence the condition for tachyonic reheating is trivially satisfied.
For a value of $g$ such as $g = 10^{-1}$, the inequality (\ref{ineq})
is violated. However, for  
sufficiently small values of $g$, there is
a transition to the region in which (\ref{amp4}) rather than
(\ref{amp5}) applies. The conclusions concerning the validity
of (\ref{cond5}), however, remain the same.

To conclude this section, we have shown that the initial stages
of reheating will proceed, for reasonable parameters of the
D3/D7 brane system, via the tachyonic instability channel.
However, once $\rho_+$ has increased beyond the value
(\ref{cond4}), the tachyonic decay will terminate. At this
point, although most of the energy is contained in the
nonlinear local fluctuations, a substantial fraction of the energy 
may remain in
the background fields. To obtain an upper bound it takes
on the time for most of the energy in the homogeneous fields to
dissipate, we neglect the effects of the nonlinear fluctuations
and analyze the subsequent dynamics which then proceeds by the
parametric resonance instability during the period when the
background fields oscillate about their minima. This is the topic
of the following section.

At this point, we must add some comments about the phase of $S$.
Neglecting the phase of $S$ means that the two branes (point particles
in the $x_4-x_5$ plane) will be approaching each other in a straight line.
This requires fine-tuned initial conditions. Also, if these initial
conditions were realized, then the D3 brane would most likely immediately
be absorbed by the D7 brane when $S = 0$ for the first time. Thus,
no oscillations of $S$ could take place. In the following, we
will be assuming that the impact parameter is sufficiently small such
that the background field dynamics for $S$ can be approximated by
an oscillation of a real-valued field $S$. Oscillations of $S$ about
$S = 0$ are interpreted as the D3 brane performing approximately
oscillatory motion about the D7 brane without zero impact parameter
being reached.  

\section{Reheating in the D3/D7 Brane Inflation Model}

As discussed in the previous section, even though the initial
stages of reheating in the D3/D7 brane inflation
model take place by tachyonic resonance, this process shuts
off early, leaving some of the initial
energy in the form of oscillations of the background fields.
To obtain an upper bound on the time it takes to drain
this residual energy from the background fields,
we proceed to study the phase when
the background fields $S$ and $\phi_+$ are oscillating
about their respective minima. Under these conditions, preheating
via a parametric resonance instability is expected to be the most
efficient mechanism for energy transfer from the inflaton to
matter field fluctuations.

We recall that for $\phi_- = 0$, the potential of our model reduces
to that of supersymmetric hybrid inflation. Reheating in
hybrid inflation was first studied in \cite{GB2}, and then
in greater depth in \cite{tachyonic2,tachyonic} where, in
particular, the tachyonic instability channel was investigated
\footnote{See also \cite{Natalia} for an application of
tachyonic decay to the cosmological moduli problem.}.
Some aspects of reheating in supersymmetric
hybrid inflation have been investigated in \cite{BKS}. In the 
following, we take up this problem again and study it from a more 
analytical point of view. 

The strength of the preheating instability depends significantly
on whether the resonance is ``broad" or ``narrow", in the notation
of \cite{KLS2}. Broad-band resonance is significantly more efficient.
However, even narrow-band resonance can be much more efficient than
the lowest order perturbative reheating channels first discussed in
\cite{DL,AFW}.
Below, we will find that, except possibly for a short period at the
beginning of the phase of oscillations, we are {\it not} in the
broad resonance region. Nevertheless, we will verify that the
resonance is sufficiently efficient such that a substantial amount
of the inflaton energy is transferred to matter fluctuations in one
Hubble expansion time. Following the philosophy first applied in
\cite{JB1},
we will neglect the expansion of space and use the efficiency 
condition just mentioned to argue that the approximation of neglecting
the spatial expansion is
self-consistent. An improved analysis could be done by going to
conformal time and rescaling the fields such that the Hubble
damping terms in the equations of motion for the rescaled fields
disappear. These equations have the form of Floquet-type
equations and show exponential instabilities. In the absence of
expansion of space, we obtain a particularly simple Floquet-type
equation, namely the Mathieu equation. 

The analyses of preheating are based on studying the equations of
motion for the field fluctuations semi-classically for a classical
background dynamics. Before turning to the discussion of the
fluctuation equations, we must specify the background evolution. 
As pointed out in \cite{BKS}, the background trajectories in the
$S-\phi_+$ plane are typically not oscillatory along a line in the
two dimensional configuration space. This introduces
complications compared to the usual studies of preheating. However, 
it was also shown in \cite{BKS} that the solutions are typically
not too far from a preferred solution to the background equations
where $S$ and $\phi_+$ oscillate in phase (i.e. along a line in
configuration space). Let us consider a particular Hubble volume
in which the fluctuation of $\phi_+$ has positive amplitude. In
this case, we introduce a rescaled field ${\tilde \phi}$ via
\be
\phi_+ \, = \, \zeta_3^{1/2} + {\tilde \phi} \, .
\ee
Since the initial value of $S$ at the end of the period of slow
rolling is approximately $S_t = \sqrt{\zeta_3 / 2}$ whereas the
amplitude of $\phi_+$ is almost vanishing, the
approximate linear background trajectory during the resonance
is characterized by
\be \label{trajec}
{\tilde \phi} \, = \, - \sqrt{2} S \, .
\ee
We observe that both fields oscillate with the same
frequency $\omega$ which is given by
\be
\omega \, = \, 2 g \zeta_3^{1/2} \, . 
\ee

The
potential of our problem is given by the sum of (\ref{pot1}) 
and (\ref{pot2}). For small values of the string coupling
constant $g$, the supersymmetry breaking term $\Delta V$ of (\ref{pot2}) 
is negligible for small values of the inflaton compared to the
first term (\ref{pot1}). Hence, for $\phi_- = 0$ the 
relevant part of the potential becomes
\be \label{potfinal}
V(S, \phi_+) \, = \, 2 g^2 S^2 \phi_+^2 
+ {{g^2} \over 2} \bigl( \phi_+^2 - \zeta_3 \bigr)^2 \, , 
\ee
where we recall that we are neglecting the phase of $\phi_+$ and thus
treating $\phi_+$ as a real variable.  

Now we are ready to derive the equations of motion for the
field fluctuations $\delta S$ and $\delta \phi$ (to simplify
the notation we are dropping the subscript $+$ on $\delta \phi$).
We take the background field
\bea \label{bgrd}
S_0 \, &=& \, {1 \over {\sqrt{2}}} {\cal A} {\rm cos}(\omega t) \\
\phi_0 \, &=& \, \zeta_3^{1/2} - {\cal A} {\rm cos}(\omega t) \, , 
\nonumber
\eea
where ${\cal A}$ is the amplitude of the background field oscillations,
and expand the potential to quadratic order in the field fluctuations
defined by
\bea
\delta S \, &=& \, S - S_0 \\
\delta \phi \, &=& \, \phi_+ - \phi_0 \, . \nonumber
\eea

Since we are expanding about a solution of the equations of motion,
the terms linear in the fluctuations in the action cancel by the
background equations of motion. The quadratic terms in the potential
are
\bea
V^{(2)}(\delta S, \delta \phi) \, &=& \, 2 g^2 \phi_0^2 {\delta S}^2
+ 8 g^2 \phi_0 S_0 \delta S \delta \phi \\
&& + 2 g^2 S_0^2 {\delta \phi}^2 
+ g^2 \bigl( 3 \phi_0^2 - \zeta_3 \bigr) {\delta \phi}^2 \, . \nonumber
\eea
Since the resulting equations of motion are linear, it is convenient
to work in momentum space.
The equations of motion for the fluctuation modes $\delta S_k$ and
$\delta \phi_k$ with comoving wavenumber $k$ hence become
\bea \label{perteom}
\ddot{\delta S_k} + \bigl( k^2 + 4 g^2 \phi_0^2 \bigr) \delta S_k \, && \\
+ 8 g^2 \phi_0 S_0 \delta \phi_k \, &=& 0 \, , \nonumber \\
\ddot{\delta \phi_k} + 
\bigl( k^2 + 4 g^2 S_0^2 + 6 g^2 \phi_0^2 - 2 g^2 \zeta_3 \bigr) \delta \phi_k
\, && \nonumber \\
+ 8 g^2 \phi_0 S_0 \delta S_k \, &=& 0 \, .
\eea

The amplitude ${\cal A}$ of the background field oscillations
starts out of the order $\zeta_3^{1/2}$ but decreases as a consequence
of the back-reaction of the fluctuations. As the fluctuations drain
energy from the background, ${\cal A}$ decreases. Since we are
interested in whether preheating is efficient at draining most of
the energy from the background, we will in the following work in the
approximation that ${\cal A}^2 \ll \zeta_3$ and neglect terms
quadratic in ${\cal A}$. Inserting the background field
expressions (\ref{bgrd}), the equation for the inflaton fluctuation
takes the form
\bea  \label{seom}
\ddot{\delta S_k} &+& \bigl( k^2 + 4 g^2 \zeta_3  
- 8 g^2 {\cal A} \zeta_3^{1/2} {\rm cos}(\omega t) \bigr) \delta S_k \nonumber \\
&=& \, - {8 \over {\sqrt{2}}} g^2 \zeta_3^{1/2} {\cal A} {\rm cos}(\omega t) 
\delta \phi_k \, .
\eea
Making use of the same approximations, the equation of motion for the matter field 
fluctuations $\delta \phi$ becomes
\bea \label{feom}
\ddot{\delta \phi_k} &+& 
\bigl( k^2 + 4 g^2 \zeta_3 - 12 g^2 {\cal A} \zeta_3^{1/2} {\rm cos}(\omega t)
\bigr) \delta \phi_k \nonumber \\
&=& \, - {8 \over {\sqrt{2}}} g^2 \zeta_3^{1/2} {\cal A} {\rm cos}(\omega t)\delta S_k
\, .
\eea
Fluctuations along the trajectory given by (\ref{trajec}) obey
\be
\delta \phi_k \, = \, - \sqrt{2} \delta S_k \, .
\ee
For these trajectories, both equations (\ref{seom}) and (\ref{feom})
take the form
\be  \label{sfeom}
\ddot{\delta S_k} + \bigl( k^2 + 4 g^2 \zeta_3  
- 16 g^2 {\cal A} \zeta_3^{1/2} {\rm cos}(\omega t) \bigr) \delta S_k \,
= \, 0 \, , 
\ee
and can be put into the standard Mathieu equation form
\be \label{Mathieu}
\chi_k^{''} + \bigl( A_k - 2 q {\rm cos}(2z) \bigr) \chi_k \, = \, 0
\ee
by introducing a rescaled time variable $z$ via
\be
2 z \, = \, \omega t \, ,
\ee
and denoting the derivative with respect to $z$ by a prime.

By comparing (\ref{seom}) and (\ref{Mathieu}) we can read off
the values of $q$ and $A_k$ for the equation of motion for
$\delta S_k$:
\bea \label{Svalues}
q \, &=& \, 8 {{\cal A} \over {\zeta_3^{1/2}}} \, , \\
A_k \, &=& \, 4 + {{k^2} \over {g^2 \zeta_3}} \, . \nonumber
\eea

The condition for broad parametric resonance is
\be
A_k \, < \, 2 q \, .
\ee
Hence, it follows from (\ref{Svalues}) that, except right
at the beginning of the reheating period when ${\cal A}^2 \simeq \zeta_3$,
the broad resonance condition is not satisfied 
\footnote{Note that the authors of \cite{BKS} did not
take into account the gravitational back-reaction which will rapidly
reduce the value of ${\cal A}$.}.

We thus conclude that the resonance is of narrow-band type. This
means \cite{LL,Arnold,KLS2} that it will take place for
values of $k$ in resonance bands centered at half integer multiples
of $\omega$. The lowest instability band is the widest. Its range is
\be
k \, \in \, {\omega \over 2}[1 - q/2, 1 + q/2] \, .
\ee

We will now show that, in spite of the fact that the resonance is
narrow rather than broad, it is strong enough to
drain a substantial fraction of the energy of the inflaton within 
one Hubble expansion time, hence justifying neglecting the expansion
of space in our analysis of the reheating equations. 

The condition for efficiency of the parametric resonance is \cite{KLS2}
\be \label{cond6}
q^2 \omega \, \gg \, H \, ,
\ee
where $H$ is the Hubble expansion rate, evaluated at the beginning of
the period of reheating. This condition comes from demanding that the
exponential growth of the fluctuations induced by the parametric
resonance instability is rapid compared to the Hubble expansion rate.
The fluctuations grow exponentially, with the exponent being 
$\mu_k t$, where $\mu_k$ is the so-called Floquet exponent given by 
\be
\mu_k \, \simeq \, {q \over 2} \, .
\ee
The condition (\ref{cond6}) comes from demanding that
$\mu_k \delta t > 1$ for the time interval $\delta t$ that a mode
initially at the center of the resonance band will remain in the resonance band
(recall that due to the Hubble expansion, modes are red-shifting with
respect to the center of the resonance bands).

Inserting the value of $q$ from (\ref{Svalues})
and the value of $H$ from (\ref{Hubble}), we find that the condition
(\ref{cond6}) for effectiveness of the parametric resonance
instability becomes
\be
\zeta_3^{1/2} \, < \, {{72} \over {\sqrt{3}}} m_{pl} \, ,
\ee
which is easily satisfied for interesting values of $\zeta_3$.

We thus conclude that, in spite of the fact that preheating occurs
in the narrow resonance region, it is sufficiently effective to
drain a substantial fraction of the energy density of the background
fields within a Hubble expansion time.

The production of matter fluctuations by the preheating instability will
drain energy from the background field oscillations and thus lead to
a decrease in the amplitude ${\cal A}$ of these oscillations. The
resonance will continue to be efficient until the condition (\ref{cond6})
ceases to be satisfied. This will be the case once $q$ decreases to
the value
\be
q_f \, = \, \Biggl({H \over {\omega}}\Biggr)^{1/2} \, ,
\ee
which corresponds to a final amplitude of
\be
{\cal A}_f^2 \, = \, {1 \over {64}}{{\zeta_3^{1/2}} \over {m_{pl}}} \zeta_3 \, .
\ee
We immediately see that, provided $\zeta_3 \ll m_{pl}^2$, a substantial
fraction $f$ of the background energy density will transfer to fluctuations
during the time period when the resonance is efficient. This fraction is
given by
\be
f \, = \, 1 - {1 \over {64}}{{\zeta_3^{1/2}} \over {m_{pl}}} \, .
\ee

We will complete this section with the discussion of two side issues.
The first is the role of the mixing terms in the set of fluctuation
equations (\ref{perteom}). In principle, since the mass matrix for
the fluctuations is symmetric, it could be diagonalized. Since the
mixing terms are of the order of ${\cal A} / \zeta_3^{1/2}$, the only
effect they have would be to change the coefficients of the two $q$ 
values slightly.
Our conclusions about narrow versus broad resonance and about the
efficiency of preheating are unchanged.

The second issue concerns possible fluctuations in the $\phi_-$ field
produced during the preheating period. In the vacuum, $\phi_- = 0$
(see Section 3), and this is the value we will expand about. Note also
that $\phi_-$ is real and positive semi-definite. One can give a plausible 
explanation using duality arguments. Under a U-duality the string connecting 
the D3 and the D7 branes map to a wrapped D3 brane on a three cycle of a 
deformed conifold \cite{BSV}. This is a massive black-hole in four dimensions
and gives rise to a unique charged 4D hypermultiplet \cite{Sb}. 
The expectation value of the scalars in this 
hypermultiplet gives the 4D black-hole configuration, and 
therefore it makes sense to take both $\phi_\pm$ to be positive definite.   
This observation removes the apparent
non-analyticity in the potential coming from the term proportional to
$|\phi_+ \phi_-|$. As a consequence, every fluctuation about $\phi_- = 0$
must be positive semi-definite in space and hence must have a non-vanishing
spatial average. This average is subject to the linear confining potential
mentioned above, and hence there is no possibility of a growth in
fluctuations of $\phi_-$.

\section{Entropy and Induced Secondary Curvature Fluctuations}

In this section we turn to the study of entropy fluctuations generated
during the initial stages of reheating, and to the calculation of the
induced curvature fluctuations. The fact that entropy fluctuations
induce a growing curvature mode has been known from the early days
of cosmological perturbation theory. An early application in the
context of inflationary cosmology is to the computation of curvature
fluctuations induced by axion perturbations in the early universe
(see e.g. \cite{Minos}). We will use the more recent formalism of
Gordon et al. \cite{Gordon} for our analysis.

We will focus on the effects of a single entropy mode and thus consider
a two field system, the initial inflaton field $S$ (which dominates the
energy-momentum tensor) and the initial entropy field $\phi_{+}$. The equation
which describes the growth of the induced curvature perturbation on
super-Hubble scales is \cite{Gordon}
\be \label{basiceq}
{\dot {{\cal R}}} \, = \, - {{2H} \over {{\dot \sigma}^2}} V_{s} \delta s \, ,
\ee
where ${\cal R}$ is the curvature fluctuation in the comoving coordinate
system, $\sigma$ is the effective adiabatic field, $s$ is the effective
entropy field, and $V_{s}$ is the derivative of the potential with respect
to the field $s$. The effective fields $\sigma$ and $s$ are combinations
of the inflaton field $S$ and tachyonic field $\phi_{+}$ which depend
on the background motion:
\be
{\dot \sigma} \, = \, {\dot S}~{\rm cos}~\theta + {\dot \phi_{+}}  ~ {\rm sin}~{\theta} 
\ee
and
\be
\delta s \, = \, \delta \phi_{+} ~{\rm cos}~\theta + \delta S ~ {\rm sin}~\theta \, , 
\ee
where the angle $\theta$ is given by
\be
{\rm cos}~\theta \, = \, {{\dot S} \over {\sqrt{ {\dot S}^2 + {\dot \phi_{+}}^2}}} 
\, .
\ee
During the early stages of the tachyonic instability, the inflaton velocity
is larger than the velocity of $\phi_{+}$. Hence, the angle $\theta$ is
approximately zero. 

The potential of our system is given by (\ref{potfinal}). Hence, the Hubble
constant $H$ at the onset of reheating is approximately given by
(\ref{Hubble}).

During the tachyonic resonance period, the derivative $V_{s}$ entering into
our basic equation (\ref{basiceq}) is approximately given by
\footnote{Note that the inflation field $S$ has non-vanishing ${\dot S}$ once
it hits the instability point $S = S_t$. It thus rolls on, and terms
in $V_{\phi_{+}}$ proportional to $S^2$ are negligible.}
\be
V_{s} \, \sim \, - g^2 \zeta_3 \phi_{+} \, .
\ee
Since the Hubble friction is negligible, the equation of motion for
the entropy field $\phi_{+}$ during this phase is
\be
{\ddot \phi_{+}} \, = \, g^2 \zeta_3 \phi_{+} \, ,
\ee
which has exponentially growing solutions
\be
\phi_{+}(t) \, = \, \phi_{+}(0) e^{\mu_F t} \, ,
\ee
with Floquet exponent $\mu_F$ given by
\be
\mu_F \, = \, g \zeta_3^{1/2} \, .
\ee
In the above, we set the time at the beginning of the instability to be
$t = 0$.

Neglecting the Hubble damping, the equation of motion for the
entropy fluctuation $\delta s \simeq \delta \phi_{+}$ on super-Hubble
scales becomes
\be
{\ddot {\delta s}} \, = \, 
\Biggl( V_{ss} + 3 {{V_s^2} \over {{\dot \sigma}^2}} \Biggr) \delta s \, ,
\ee
where the subscripts on the potential indicate which fields the derivative
is taken with respect to. While the tachyonic instability condition
(\ref{cond4}) is satisfied, this equation can be approximated by
\be
{\ddot {\delta s}} \, = \, g^2 \zeta_3 \delta s \, .
\ee
This shows that the entropy fluctuation increases with the
same Floquet exponent as the background entropy field:
\be
\delta s (t) \, = \, \delta s (0) e^{\mu_F t} \, .
\ee
The initial values of $\delta s$ and $\phi_{+}$ are both given
by the same large scale quantum fluctuations and will be taken
to be of the same order of magnitude.

Having determined the growth of the entropy mode, we can now
integrate (\ref{basiceq}) from the time $t = 0$ when the
instability starts to the time when the tachyonic instability
condition breaks down.     We denote the corresponding time 
by $t = t_f$. In the absence of back-reaction, the
tachyonic resonance stops when (\ref{cond4}) is satisfied.
We denote the result of the integration
by $\Delta {\cal R}$. An approximate evaluation of the integral
(taking ${\dot \sigma}$ to be constant) gives
\be
\Delta {\cal R}(t_f) \, = \, g {{1} \over {2 m_{pl} \mu_F}} 
\phi_{+}(0) \delta s (0) e^{2 \mu_F t_f} \, .
\ee
  In the absence of back-reaction, then from (\ref{cond4})
\be
\phi_{+}(t_f) \, = \, e^{\mu_F t_f} \phi_{+}(0) \, \sim \, \zeta_3^{1/2}
\ee
and analogously for $\delta s(t_f)$, we can solve for the duration $t_f$
of the tachyonic instability. We see that the dependence of our result
for $\Delta {\cal R}$ on the initial value of the entropy mode drops
out \footnote{A similar cancellation was seen in the work \cite{Larissa}.}
and we obtain   the result
\be \label{finalresult}
\Delta {\cal R} \, \sim \, {{\zeta_3^{1/2}} \over {m_{pl}}} \, .
\ee

    Let us now turn to a brief discussion of back-reaction effects. The
"waterfall" field $\phi_{+}$ has a dispersion $\sigma$ on microscopic
scales which are due to its quantum vacuum fluctuations. By integrating
up these quantum fluctuations of $\phi_{+}$ to its mass scale $m$,
we obtain the following result for the initial dispersion
$D(0)$ (that is the root mean square value of the field) 
at the time the tachyonic instability sets in
\be
D(0) \, \sim \, m \, .
\ee
This dispersion then grows exponentially with the exponent set by
the Floquet exponent $\mu_F$, and after a time
\be
t_s \, \simeq \, \mu_F {\rm ln}\left({{\zeta_3} \over {D(0)}}\right)
\ee
the dispersion will have grown to be comparable to the value of
$\phi_{+}$ at the minimum of the potential. At this time (called
the spinodal decomposition time), the field on small scales fragments
into domains of typical size $m^{-1}$. While in itself the
formation of nonlinearities on small distance scales does not
interfere with the linear growth of fluctuations on cosmological
scales, the nonlinearities can induce a positive contribution
to the effective square mass of the tachyon field which shuts
off the resonance. 

Whether back-reaction shuts off the resonance before the
entropy fluctuation has had time to fully develop depends a lot
on the values of the dispersion $D(0)$ and of
the quasi-homogeneous mode $\phi_{+}$ on the
scale $k$ of the fluctuation. If the latter is calculated based on
energetics, i.e. by setting the energy density in this mode to be
comparable to the quantum vacuum energy density
\be
\phi_{+}^2 m^2 \, \sim \, H^4  \, ,
\ee
then the time scales $t_s$ and $t_f$ are comparable and our
estimates hold. However, if $\phi_{+}$ is estimated by
integrating quantum vacuum fluctuations on length scales larger 
than $k^{-1}$, and taking into account that the waterfall field
is massive during most of the inflationary period and the
amplitude of the fluctuations is hence redshifted, then $t_s$
is much smaller than $t_f$, and hence the growth of the
entropy fluctuations is cut off before they can become significant.

  Let us for a moment assume that $t_f$ and $t_s$ are comparable. In
this case, our
result (\ref{finalresult}) for the amplitude of the secondary
curvature fluctuations induced by the entropy mode must be compared
to the amplitude of the primary adiabatic mode. In the parameter
regime where (\ref{ineq}) is satisfied, i.e. 
\be
\zeta_3^{1/2} \, > \, g m_{pl} \, ,
\ee
the primary adiabatic fluctuations are given by (see (\ref{amp4}))
\be
{\cal R}_{\rm adiab} \, \sim \, g^{-1} \zeta_3^{3/2} m_{pl}^{-3} \, .
\ee
Hence, in the parameter region given by
\be
g m_{pl} \, < \, \zeta_3^{1/2} \, <  \, g^{1/2} m_{pl} 
\ee
the secondary fluctuations dominate over the primary ones.

When the condition (\ref{ineq}) is not satisfied, the amplitude of
the primary adiabatic modes is given by (see (\ref{amp5}))
\be
{\cal R}_{\rm adiab} \, \sim \, \sqrt{N} {{\zeta_3} \over {m_{pl}^2}} \, ,
\ee
and hence the secondary curvature fluctuations dominate over the
primary ones unless
\be
\sqrt{N} \zeta_3^{1/2} \, > \, m_{pl} \, .
\ee

From the above results we can draw two main conclusions. Firstly, we
see that the induced curvature fluctuation produced by the entropy
mode remains in the linear regime. This is good news for the model.
The second conclusion is that   
it is possible the secondary fluctuations dominate over the primary ones, thus
necessitating a change in the model parameters in order to achieve
agreement with the observed amplitude of fluctuations on large scales.

\section{Discussion and Conclusions}

We have studied reheating and structure formation in the D3/D7 brane 
inflation model of \cite{DHHK}, with particular emphasis on the
tachyonic instability of super-Hubble scale entropy fluctuations.
These entropy perturbations induce a curvature fluctuations which we
call ``secondary''. We have found that   under certain conditions 
these secondary fluctuations are larger
than the primordial adiabatic ones which have been considered in
the past. They do, however, remain in the regime of applicability of
linear perturbation theory.   This would imply that the parameters
of the model have to be changed compared to what is usually assumed
in order to obtain agreement with the observed amplitude of the
large-scale curvature fluctuations. 

Along the way, we have given an extensive analytical analysis of
the reheating mechanism in the D3/D7 brane inflation 
model\footnote{Our analytical analysis is complementary to the numerical work
of \cite{tachyonic}. We need to make certain approximations and neglect
some effects, as discussed in the text. On the other hand, numerical work
can only handle a limited range of scales. We are interested in large
cosmological scales, whereas the basic physical scale of the system is
microphysical. It is not clear that numerical work which is sensitive
to the microphysical scale can make reliable predictions for questions
involving cosmological scales.}.  
In the low energy field theory limit, the
dynamics of this system is a special case of supersymmetric hybrid
inflation. We have taken into account the quantum fluctuations in the
``waterfall'' field $\phi_+$. These play an important role for both
the reheating dynamics and for the generation of entropy
fluctuations. For COBE-normalized values of the string theory parameters
and for reasonable values of the string coupling constant we have
found that, including the effects of the above mentioned quantum fluctuations,
the initial stages of reheating occur via a tachyonic instability.
This instability, however, shuts off quite early, leaving 
some of the initial inflaton energy in the
background fields. To obtain an upper bound on the time it takes
to drain most of the residual energy from the background homogeneous
fields, we have neglected the interactions of the nonlinear fluctuations
produced during the initial phase of tachyonic decay, and focused
on the residual homogeneous field dynamics. This later dynamics proceeds
by the parametric resonance instability of \cite{JB1,KLS1,JB2,KLS2}.
In fact, except for at the onset of the reheating process, the system
is in the narrow-band region of parameters. Nevertheless, the reheating
is sufficiently efficient to convert a substantial fraction of the
inflaton energy into matter fluctuations. 

We note that efficient reheating in the D3/D7 brane inflation model is 
easier to achieve than in multi-throat inflation models since the matter
fields are directly coupled to the inflaton in the form of strings
stretching between the two branes whose separation constitutes the
inflaton.

In our work, we have neglected the issue of moduli stabilization.
This is a very important caveat to our analysis. Moduli stabilization
in our model has been considered in the first reference of
\cite{previous} and more recently in
\cite{BBDD}. We plan to study the implications of the corrections to
the potential induced by moduli stabilization on the dynamics of
reheating in a followup paper.

We have also simplified the dynamics of the background fields. Namely,
we have considered in-phase oscillations of the two background fields
$S$ and $\phi_+$, and we have neglected their phases. These phases
provide extra low mass entropy modes. It would be interesting to
consider their excitation during reheating.

\begin{acknowledgments}

The work of R.B. and K.D. is supported by funds from McGill 
University, by NSERC Discovery Grants and by the Canada Research Chairs 
program. The work of ACD is supported in part by PPARC. She wishes
to thank the Physics Department, McGill University for hospitality
whilst this work was in progress. We are grateful to Neil Barnaby,
Jim Cline and Andrew Frey for discussions, and to Renata Kallosh,
Lev Kofman and Andrei Linde for comments on an earlier draft. We also thank the 
referee for his/her comments that helped us to improve the paper.

\end{acknowledgments}


\begin{thebibliography}{99}

\bibitem{RHBrev}
R.~H.~Brandenberger,  
   ``Inflationary cosmology: Progress and problems,''  
   arXiv:hep-ph/9910410.  
   %%CITATION = HEP-PH 9910410;%%

\bibitem{Linderev}
A.~Linde,  
   ``Inflation and string cosmology,''
   eConf {\bf C040802}, L024 (2004)  
   [J.\ Phys.\ Conf.\ Ser.\  {\bf 24}, 151 (2005)]  
   [arXiv:hep-th/0503195].  
   %%CITATION = HEP-TH 0503195;%%

\bibitem{Cliffrev}
C.~P.~Burgess,  
   ``Inflatable string theory?,''  
   Pramana {\bf 63}, 1269 (2004)  
   [arXiv:hep-th/0408037].  
   %%CITATION = HEP-TH 0408037;%%

\bibitem{JCrev}
J.~M.~Cline,  
   ``Inflation from string theory,''
   arXiv:hep-th/0501179.
   %%CITATION = HEP-TH 0501179;%%

\bibitem{EvaLiam}
L.~McAllister and E.~Silverstein,
  ``String Cosmology: A Review,''
  arXiv:0710.2951 [hep-th].
  %%CITATION = ARXIV:0710.2951;%%

\bibitem{Dvali}
G.~R.~Dvali and S.~H.~H.~Tye,
  ``Brane inflation,''  
  Phys.\ Lett.\ B {\bf 450}, 72 (1999)
  [arXiv:hep-ph/9812483].
  %%CITATION = HEP-PH 9812483;%%

\bibitem{Stephon}
S.~H.~S.~Alexander,
  ``Inflation from D - anti-D brane annihilation,''
  Phys.\ Rev.\ D {\bf 65}, 023507 (2002)
  [arXiv:hep-th/0105032].
  %%CITATION = HEP-TH 0105032;%%

\bibitem{Shafi}
G.~R.~Dvali, Q.~Shafi and S.~Solganik,
  ``D-brane inflation,''
  arXiv:hep-th/0105203.
  %%CITATION = HEP-TH 0105203;%%

\bibitem{Cliff}
C.~P.~Burgess, M.~Majumdar, D.~Nolte, F.~Quevedo, G.~Rajesh and R.~J.~Zhang,
  ``The inflationary brane-antibrane universe,''
  JHEP {\bf 0107}, 047 (2001)
  [arXiv:hep-th/0105204].
  %%CITATION = HEP-TH 0105204;%%

\bibitem{JuanGB}
J.~Garcia-Bellido, R.~Rabadan and F.~Zamora,
  ``Inflationary scenarios from branes at angles,''
  JHEP {\bf 0201}, 036 (2002)
  [arXiv:hep-th/0112147];\\
  %%CITATION = HEP-TH 0112147;%%
N.~T.~Jones, H.~Stoica and S.~H.~H.~Tye,
  ``Brane interaction as the origin of inflation,''
  JHEP {\bf 0207}, 051 (2002)
  [arXiv:hep-th/0203163];\\
  %%CITATION = HEP-TH 0203163;%%
M.~Gomez-Reino and I.~Zavala,
  ``Recombination of intersecting D-branes and cosmological inflation,''
  JHEP {\bf 0209}, 020 (2002)
  [arXiv:hep-th/0207278].
  %%CITATION = HEP-TH 0207278;%%

\bibitem{KKLMMT}
S.~Kachru, R.~Kallosh, A.~Linde, J.~M.~Maldacena, L.~McAllister and S.~P.~Trivedi,
  ``Towards inflation in string theory,''
  JCAP {\bf 0310}, 013 (2003)
  [arXiv:hep-th/0308055].
  %%CITATION = HEP-TH 0308055;%%

\bibitem{DHHK}
K.~Dasgupta, C.~Herdeiro, S.~Hirano and R.~Kallosh,
  ``D3/D7 inflationary model and M-theory,''
  Phys.\ Rev.\ D {\bf 65}, 126002 (2002)
  [arXiv:hep-th/0203019].
  %%CITATION = HEP-TH 0203019;%%

\bibitem{shift}
M.~Kawasaki, M.~Yamaguchi and T.~Yanagida,
  ``Natural chaotic inflation in supergravity and leptogenesis,''
  Phys.\ Rev.\ D {\bf 63}, 103514 (2001)
  [arXiv:hep-ph/0011104];\\
  %%CITATION = HEP-PH 0011104;%%
M.~Yamaguchi and J.~Yokoyama,
  ``New inflation in supergravity with a chaotic initial condition,''
  Phys.\ Rev.\ D {\bf 63}, 043506 (2001)
  [arXiv:hep-ph/0007021];\\
  %%CITATION = HEP-PH 0007021;%%
J.~P.~Hsu and R.~Kallosh,
  ``Volume stabilization and the origin of the inflaton shift symmetry in
  string theory,''
  JHEP {\bf 0404}, 042 (2004)
  [arXiv:hep-th/0402047].
  %%CITATION = HEP-TH 0402047;%%

\bibitem{kors}
  M.~Berg, M.~Haack and B.~Kors,
  ``Loop corrections to volume moduli and inflation in string theory,''
  Phys.\ Rev.\  D {\bf 71}, 026005 (2005)
  [arXiv:hep-th/0404087];\\
  %%CITATION = PHRVA,D71,026005;%%
  L.~McAllister,
  ``An inflaton mass problem in string inflation from threshold corrections  to
  volume stabilization,''
  JCAP {\bf 0602} (2006) 010
  [arXiv:hep-th/0502001];\\
  %%CITATION = JCAPA,0602,010;%%
  M.~Haack, R.~Kallosh, A.~Krause, A.~Linde, D.~Lust and M.~Zagermann,
  ``Update of D3/D7-Brane Inflation on $K3 \times T^2/Z_2$,'' 
  arXiv:0804.3961 [hep-th].
  %%CITATION = ARXIV:0804.3961;%%

\bibitem{herd}
  C.~Herdeiro, S.~Hirano and R.~Kallosh,
  ``String theory and hybrid inflation / acceleration,''
  JHEP {\bf 0112}, 027 (2001)
  [arXiv:hep-th/0110271].
  %%CITATION = JHEPA,0112,027;%%

\bibitem{D37strings}
K.~Dasgupta, J.~P.~Hsu, R.~Kallosh, A.~Linde and M.~Zagermann,
  ``D3/D7 brane inflation and semilocal strings,''
  JHEP {\bf 0408}, 030 (2004)
  [arXiv:hep-th/0405247].
  %%CITATION = HEP-TH 0405247;%%

\bibitem{previous}
J.~P.~Hsu, R.~Kallosh and S.~Prokushkin,
  ``On brane inflation with volume stabilization,''
  JCAP {\bf 0312}, 009 (2003)
  [arXiv:hep-th/0311077];\\
  %%CITATION = HEP-TH 0311077;%%
P.~Chen, K.~Dasgupta, K.~Narayan, M.~Shmakova and M.~Zagermann,
  ``Brane inflation, solitons and cosmological solutions: I,''
  JHEP {\bf 0509}, 009 (2005)
  [arXiv:hep-th/0501185];\\
  %%CITATION = HEP-TH 0501185;%%
F.~Koyama, Y.~Tachikawa and T.~Watari,
  ``Supergravity analysis of hybrid inflation model from D3-D7 system,''
  Phys.\ Rev.\ D {\bf 69}, 106001 (2004)
  [Erratum-ibid.\ D {\bf 70}, 129907 (2004)]
  [arXiv:hep-th/0311191].
  %%CITATION = HEP-TH 0311191;%%

\bibitem{GVW}
  S.~Gukov, C.~Vafa and E.~Witten,
  ``CFT's from Calabi-Yau four-folds,''
  Nucl.\ Phys.\ B {\bf 584}, 69 (2000)
  [Erratum-ibid.\ B {\bf 608}, 477 (2001)]
  [arXiv:hep-th/9906070].
  %%CITATION = HEP-TH 9906070;%%

\bibitem{DRS}
  K.~Dasgupta, G.~Rajesh and S.~Sethi,
  ``M theory, orientifolds and G-flux,''
  JHEP {\bf 9908}, 023 (1999)
  [arXiv:hep-th/9908088].
  %%CITATION = HEP-TH 9908088;%%

\bibitem{GKP}
  S.~B.~Giddings, S.~Kachru and J.~Polchinski,
  ``Hierarchies from fluxes in string compactifications,''
  Phys.\ Rev.\ D {\bf 66}, 106006 (2002)
  [arXiv:hep-th/0105097].
  %%CITATION = HEP-TH 0105097;%%

\bibitem{Mollerach}
S.~Mollerach,
  ``Isocurvature Baryon Perturbations And Inflation,''
  Phys.\ Rev.\  D {\bf 42}, 313 (1990).
  %%CITATION = PHRVA,D42,313;%%

\bibitem{SY}
A.~A.~Starobinsky and J.~Yokoyama,
  ``Density fluctuations in Brans-Dicke inflation,''
  arXiv:gr-qc/9502002.
  %%CITATION = GR-QC/9502002;%%

\bibitem{LM}
A.~D.~Linde and V.~F.~Mukhanov,
  ``Nongaussian isocurvature perturbations from inflation,''
  Phys.\ Rev.\  D {\bf 56}, 535 (1997)
  [arXiv:astro-ph/9610219].
  %%CITATION = PHRVA,D56,535;%%

\bibitem{Moroi}
T.~Moroi and T.~Takahashi,
  ``Effects of cosmological moduli fields on cosmic microwave background,''
  Phys.\ Lett.\  B {\bf 522}, 215 (2001)
  [Erratum-ibid.\  B {\bf 539}, 303 (2002)]
  [arXiv:hep-ph/0110096];\\
  %%CITATION = PHLTA,B522,215;%%
T.~Moroi and T.~Takahashi,
  ``Cosmic density perturbations from late-decaying scalar condensations,''
  Phys.\ Rev.\  D {\bf 66}, 063501 (2002)
  [arXiv:hep-ph/0206026].
  %%CITATION = PHRVA,D66,063501;%%

\bibitem{LW}
D.~H.~Lyth and D.~Wands,
  ``Generating the curvature perturbation without an inflaton,''
  Phys.\ Lett.\  B {\bf 524}, 5 (2002)
  [arXiv:hep-ph/0110002].
  %%CITATION = PHLTA,B524,5;%%

\bibitem{Sloth}
K.~Enqvist and M.~S.~Sloth,
  ``Adiabatic CMB perturbations in pre big bang string cosmology,''
  Nucl.\ Phys.\  B {\bf 626}, 395 (2002)
  [arXiv:hep-ph/0109214].
  %%CITATION = NUPHA,B626,395;%%

\bibitem{DGZ}
G.~Dvali, A.~Gruzinov and M.~Zaldarriaga,
  ``A new mechanism for generating density perturbations from inflation,''
  Phys.\ Rev.\  D {\bf 69}, 023505 (2004)
  [arXiv:astro-ph/0303591].
  %%CITATION = PHRVA,D69,023505;%%

\bibitem{Lev}
L.~Kofman,
  ``Probing string theory with modulated cosmological fluctuations,''
  arXiv:astro-ph/0303614.
  %%CITATION = ASTRO-PH/0303614;%%

\bibitem{MR}
S.~Matarrese and A.~Riotto,
  ``Large-scale curvature perturbations with spatial and time variations of
  the inflaton decay rate,''
  JCAP {\bf 0308}, 007 (2003)
  [arXiv:astro-ph/0306416].
  %%CITATION = JCAPA,0308,007;%%

\bibitem{Uzan}
F.~Bernardeau, L.~Kofman and J.~P.~Uzan,
  ``Modulated fluctuations from hybrid inflation,''
  Phys.\ Rev.\  D {\bf 70}, 083004 (2004)
  [arXiv:astro-ph/0403315].
  %%CITATION = PHRVA,D70,083004;%%

\bibitem{Vernizzi}
F.~Vernizzi,
  ``Cosmological perturbations from varying masses and couplings,''
  Phys.\ Rev.\  D {\bf 69}, 083526 (2004)
  [arXiv:astro-ph/0311167].
  %%CITATION = PHRVA,D69,083526;%%

\bibitem{KRV}
E.~W.~Kolb, A.~Riotto and A.~Vallinotto,
  ``Curvature perturbations from broken symmetries,''
  Phys.\ Rev.\  D {\bf 71}, 043513 (2005)
  [arXiv:astro-ph/0410546].
  %%CITATION = PHRVA,D71,043513;%%

\bibitem{Matsuda}
T.~Matsuda,
  ``Generating the curvature perturbation with instant preheating,''
  JCAP {\bf 0703}, 003 (2007)
  [arXiv:hep-th/0610232].
  %%CITATION = JCAPA,0703,003;%%

\bibitem{BaVi}
B.~A.~Bassett and F.~Viniegra,
  ``Massless metric preheating,''
  Phys.\ Rev.\  D {\bf 62}, 043507 (2000)
  [arXiv:hep-ph/9909353].
  %%CITATION = PHRVA,D62,043507;%%

\bibitem{FB2}
F.~Finelli and R.~H.~Brandenberger,
  ``Parametric amplification of metric fluctuations during reheating in two
  field models,''
  Phys.\ Rev.\  D {\bf 62}, 083502 (2000)
  [arXiv:hep-ph/0003172].
  %%CITATION = PHRVA,D62,083502;%%

\bibitem{FB1}
F.~Finelli and R.~H.~Brandenberger,
  ``Parametric amplification of gravitational fluctuations during  reheating,''
  Phys.\ Rev.\ Lett.\  {\bf 82}, 1362 (1999)
  [arXiv:hep-ph/9809490].
  %%CITATION = PRLTA,82,1362;%%

\bibitem{Zibin}
J.~P.~Zibin, R.~H.~Brandenberger and D.~Scott,
  ``Backreaction and the parametric resonance of cosmological  fluctuations,''
  Phys.\ Rev.\  D {\bf 63}, 043511 (2001)
  [arXiv:hep-ph/0007219].
  %%CITATION = PHRVA,D63,043511;%%

\bibitem{JB1}
J.~H.~Traschen and R.~H.~Brandenberger,
  ``Particle Production During Out-Of-Equilibrium Phase Transitions,''
  Phys.\ Rev.\ D {\bf 42}, 2491 (1990).
  %%CITATION = PHRVA,D42,2491;%%

\bibitem{KLS1}
L.~Kofman, A.~D.~Linde and A.~A.~Starobinsky,
  ``Reheating after inflation,''
  Phys.\ Rev.\ Lett.\  {\bf 73}, 3195 (1994)
  [arXiv:hep-th/9405187].
  %%CITATION = HEP-TH 9405187;%%

\bibitem{JB2}
Y.~Shtanov, J.~H.~Traschen and R.~H.~Brandenberger,
  ``Universe reheating after inflation,''
  Phys.\ Rev.\ D {\bf 51}, 5438 (1995)
  [arXiv:hep-ph/9407247].
  %%CITATION = HEP-PH 9407247;%%

\bibitem{KLS2}
L.~Kofman, A.~D.~Linde and A.~A.~Starobinsky,
  ``Towards the theory of reheating after inflation,''
  Phys.\ Rev.\ D {\bf 56}, 3258 (1997)
  [arXiv:hep-ph/9704452].
  %%CITATION = HEP-PH 9704452;%%

\bibitem{Larissa}
R.~H.~Brandenberger, A.~R.~Frey and L.~C.~Lorenz,
  ``Entropy Fluctuations in Brane Inflation Models,''
  arXiv:0712.2178 [hep-th].
  %%CITATION = ARXIV:0712.2178;%%

\bibitem{BC2}
N.~Barnaby and J.~M.~Cline,
  ``Nongaussian and nonscale-invariant perturbations from tachyonic  preheating
  in hybrid inflation,''
  Phys.\ Rev.\  D {\bf 73}, 106012 (2006)
  [arXiv:astro-ph/0601481];\\
  %%CITATION = PHRVA,D73,106012;%%
N.~Barnaby and J.~M.~Cline,
  ``Nongaussianity from tachyonic preheating in hybrid inflation,''
  Phys.\ Rev.\  D {\bf 75}, 086004 (2007)
  [arXiv:astro-ph/0611750].
  %%CITATION = PHRVA,D75,086004;%%

\bibitem{Losic}
B.~Losic and W.~G.~Unruh,
  ``Long-wavelength metric backreactions in slow-roll inflation,''
  Phys.\ Rev.\  D {\bf 72}, 123510 (2005)
  [arXiv:gr-qc/0510078].
  %%CITATION = PHRVA,D72,123510;%%

\bibitem{Patrick} 
P.~Martineau and R.~Brandenberger,
  ``A Back-reaction Induced Lower Bound on the Tensor-to-Scalar Ratio,''
  arXiv:0709.2671 [astro-ph].
  %%CITATION = ARXIV:0709.2671;%%

\bibitem{tachyonic}
G.~N.~Felder, L.~Kofman and A.~D.~Linde,
  ``Tachyonic instability and dynamics of spontaneous symmetry breaking,''
  Phys.\ Rev.\ D {\bf 64}, 123517 (2001)
  [arXiv:hep-th/0106179].
  %%CITATION = HEP-TH 0106179;%%

\bibitem{beasly}
  C.~Beasley, J.~J.~Heckman and C.~Vafa,
  ``GUTs and Exceptional Branes in F-theory - I,''
  arXiv:0802.3391 [hep-th];
`GUTs and Exceptional Branes in F-theory - II: Experimental Predictions,''
  arXiv:0806.0102 [hep-th].
  %%CITATION = ARXIV:0806.0102;%%
  %%CITATION = ARXIV:0802.3391;%%

\bibitem{Easson}
J.~H.~Brodie and D.~A.~Easson,
  ``Brane inflation and reheating,''
  JCAP {\bf 0312}, 004 (2003)
  [arXiv:hep-th/0301138].
  %%CITATION = HEP-TH 0301138;%%

\bibitem{CFM}
J.~M.~Cline, H.~Firouzjahi and P.~Martineau,
  ``Reheating from tachyon condensation,''
  JHEP {\bf 0211}, 041 (2002)
  [arXiv:hep-th/0207156].
  %%CITATION = HEP-TH 0207156;%%

\bibitem{BC}
N.~Barnaby and J.~M.~Cline,
  ``Creating the universe from brane-antibrane annihilation,''
  Phys.\ Rev.\ D {\bf 70}, 023506 (2004)
  [arXiv:hep-th/0403223].
  %%CITATION = HEP-TH 0403223;%%

\bibitem{TM}
Y.~i.~Takamizu and K.~i.~Maeda,
  ``Collision of domain walls and reheating of the brane universe,''
  Phys.\ Rev.\ D {\bf 70}, 123514 (2004)
  [arXiv:hep-th/0406235].
  %%CITATION = HEP-TH 0406235;%%

\bibitem{BBC}
N.~Barnaby, C.~P.~Burgess and J.~M.~Cline,
  ``Warped reheating in brane-antibrane inflation,''
  JCAP {\bf 0504}, 007 (2005)
  [arXiv:hep-th/0412040].
  %%CITATION = HEP-TH 0412040;%%

\bibitem{KY}
L.~Kofman and P.~Yi,
  ``Reheating the universe after string theory inflation,''
  Phys.\ Rev.\ D {\bf 72}, 106001 (2005)
  [arXiv:hep-th/0507257].
  %%CITATION = HEP-TH 0507257;%%

\bibitem{DSU}
D.~Chialva, G.~Shiu and B.~Underwood,
  ``Warped reheating in multi-throat brane inflation,''
  JHEP {\bf 0601}, 014 (2006)
  [arXiv:hep-th/0508229].
  %%CITATION = HEP-TH 0508229;%%

\bibitem{FMM}
A.~R.~Frey, A.~Mazumdar and R.~Myers,
  ``Stringy effects during inflation and reheating,''
  Phys.\ Rev.\ D {\bf 73}, 026003 (2006)
  [arXiv:hep-th/0508139].
  %%CITATION = HEP-TH 0508139;%%

\bibitem{CT}
X.~Chen and S.~H.~Tye,
  ``Heating in brane inflation and hidden dark matter,''
  [arXiv:hep-th/0602136].
  %%CITATION = HEP-TH 0602136;%%

\bibitem{PL}
P.~Langfelder,
  ``On tunnelling In two-throat warped reheating,''
  [arXiv:hep-th/0602296].
  %%CITATION = HEP-TH 0602296;%%

\bibitem{instanton}
N.~Nekrasov and A.~S.~Schwarz,
  ``Instantons on noncommutative R**4 and (2,0) superconformal six  dimensional
  theory,''
  Commun.\ Math.\ Phys.\  {\bf 198}, 689 (1998)
  [arXiv:hep-th/9802068];\\
%%CITATION = HEP-TH 9802068;%%
N.~Seiberg and E.~Witten,
  ``String theory and noncommutative geometry,''
  JHEP {\bf 9909}, 032 (1999)
  [arXiv:hep-th/9908142].
  %%CITATION = HEP-TH 9908142;%%

\bibitem{DiS}
M.~Dine and N.~Seiberg,
  ``Couplings And Scales In Superstring Models,''
  Phys.\ Rev.\ Lett.\  {\bf 55}, 366 (1985).
  %%CITATION = PRLTA,55,366;%%

\bibitem{DFKS}
K.~Dasgupta, P.~Franche, A.~Knauf and J.~Sully,
  ``D-terms on the resolved conifold,''
  arXiv:0802.0202 [hep-th].
  %%CITATION = ARXIV:0802.0202;%%

\bibitem{witten85}
  E.~Witten,
  ``Dimensional Reduction Of Superstring Models,''
  Phys.\ Lett.\  B {\bf 155}, 151 (1985).
  %%CITATION = PHLTA,B155,151;%%

\bibitem{BCDF}
  C.~P.~Burgess, J.~M.~Cline, K.~Dasgupta and H.~Firouzjahi,
  ``Uplifting and inflation with D3 branes,''
  JHEP {\bf 0703}, 027 (2007)
  [arXiv:hep-th/0610320].
  %%CITATION = JHEPA,0703,027;%%

\bibitem{mcall}
  D.~Baumann, A.~Dymarsky, I.~R.~Klebanov and L.~McAllister,
  ``Towards an Explicit Model of D-brane Inflation,''
  arXiv:0706.0360 [hep-th].
  %%CITATION = ARXIV:0706.0360;%%

\bibitem{krause}
  A.~Krause and E.~Pajer,
  ``Chasing Brane Inflation in String-Theory,''
  arXiv:0705.4682 [hep-th].
  %%CITATION = ARXIV:0705.4682;%%

\bibitem{KKLT}
  S.~Kachru, R.~Kallosh, A.~Linde and S.~P.~Trivedi,
  ``De Sitter vacua in string theory,''
  Phys.\ Rev.\ D {\bf 68}, 046005 (2003)
  [arXiv:hep-th/0301240].
  %%CITATION = HEP-TH 0301240;%%

\bibitem{DDFGK}
  F.~Denef, M.~R.~Douglas, B.~Florea, A.~Grassi and S.~Kachru,
  ``Fixing all moduli in a simple F-theory compactification,''
  [arXiv:hep-th/0503124].
  %%CITATION = HEP-TH 0503124;%%

\bibitem{Brax}
Ph.~Brax, C.~van de Bruck, A.~C.~Davis, S.~C.~Davis, 
R.~Jeannerot and M.~Postma,
  ``Moduli corrections to D-term inflation,''
  JCAP {\bf 0701}, 026 (2007)
  [arXiv:hep-th/0610195].
  %%CITATION = JCAPA,0701,026;%%

\bibitem{lumps}
  K.~Dasgupta, H.~Firouzjahi and R.~Gwyn,
  ``Lumps in the throat,''
  JHEP {\bf 0704}, 093 (2007)
  [arXiv:hep-th/0702193].
  %%CITATION = JHEPA,0704,093;%%

\bibitem{AV}
  T.~Vachaspati and A.~Achucarro,
  ``Semilocal cosmic strings,''
  Phys.\ Rev.\ D {\bf 44}, 3067 (1991).
  %%CITATION = PHRVA,D44,3067;%%

\bibitem{UAD}
  J.~Urrestilla, A.~Achucarro and A.~C.~Davis,
  ``D-term inflation without cosmic strings,''
  Phys.\ Rev.\ Lett.\  {\bf 92}, 251302 (2004)
  [arXiv:hep-th/0402032].
  %%CITATION = HEP-TH 0402032;%%

\bibitem{Burrage}
C.~Burrage and A.~C.~Davis,
  ``P-term potentials from 4-D supergravity,''
  JHEP {\bf 0706}, 086 (2007)
  [arXiv:0705.1657 [hep-th]];\\
  %%CITATION = JHEPA,0706,086;%%
C.~Burrage and A.~C.~Davis,
  ``P-term Strings and Semi-local Strings,''
  JHEP {\bf 0711}, 023 (2007)
  [arXiv:0707.3610 [hep-th]].
  %%CITATION = JHEPA,0711,023;%%

\bibitem{Jon}
J.~Urrestilla, N.~Bevis, M.~Hindmarsh, M.~Kunz and A.~R.~Liddle,
  ``Cosmic microwave anisotropies from BPS semilocal strings,''
  arXiv:0711.1842 [astro-ph].
  %%CITATION = ARXIV:0711.1842;%%

\bibitem{Dterm}
E.~D.~Stewart,
  ``Inflation, supergravity and superstrings,''
  Phys.\ Rev.\ D {\bf 51}, 6847 (1995)
  [arXiv:hep-ph/9405389];\\
  %%CITATION = HEP-PH 9405389;%%
P.~Binetruy and G.~R.~Dvali,
  ``D-term inflation,''
  Phys.\ Lett.\ B {\bf 388}, 241 (1996)
  [arXiv:hep-ph/9606342];\\
  %%CITATION = HEP-PH 9606342;%%
E.~Halyo,
  ``Hybrid inflation from supergravity D-terms,''
  Phys.\ Lett.\ B {\bf 387}, 43 (1996)
  [arXiv:hep-ph/9606423].
  %%CITATION = HEP-PH 9606423;%%

\bibitem{hybrid}
A.~D.~Linde,
  ``Hybrid inflation,''
  Phys.\ Rev.\ D {\bf 49}, 748 (1994)
  [arXiv:astro-ph/9307002].
  %%CITATION = ASTRO-PH 9307002;%%

\bibitem{Pterm}
R.~Kallosh and A.~Linde,
  ``P-term, D-term and F-term inflation,''
  JCAP {\bf 0310}, 008 (2003)
  [arXiv:hep-th/0306058].
  %%CITATION = HEP-TH 0306058;%%

\bibitem{Renata}
R.~Kallosh,
  ``N = 2 supersymmetry and de Sitter space,''
  arXiv:hep-th/0109168.
  %%CITATION = HEP-TH 0109168;%%

\bibitem{MFB}
V.~F.~Mukhanov, H.~A.~Feldman and R.~H.~Brandenberger,
  ``Theory Of Cosmological Perturbations. Part 1. Classical Perturbations. Part
  2. Quantum Theory Of Perturbations. Part 3. Extensions,''
  Phys.\ Rept.\  {\bf 215}, 203 (1992).
  %%CITATION = PRPLC,215,203;%%

\bibitem{LR}
D.~H.~Lyth and A.~Riotto,
  ``Particle physics models of inflation and the cosmological density
  perturbation,''
  Phys.\ Rept.\  {\bf 314}, 1 (1999)
  [arXiv:hep-ph/9807278].
  %%CITATION = HEP-PH 9807278;%%

\bibitem{RHBrev2}
R.~H.~Brandenberger,
  ``Lectures on the theory of cosmological perturbations,''
  Lect.\ Notes Phys.\  {\bf 646}, 127 (2004)
  [arXiv:hep-th/0306071].
  %%CITATION = HEP-TH 0306071;%%

\bibitem{Rachel}
  R.~Jeannerot and M.~Postma,
  ``Enlarging the parameter space of standard hybrid inflation,''
  JCAP {\bf 0607} (2006) 012
  [arXiv:hep-th/0604216].
  %%CITATION = JCAPA,0607,012;%%

\bibitem{Starob}
A.~A.~Starobinsky,
  ``Stochastic De Sitter (Inflationary) Stage In The Early Universe,''
  in *De Vega, H.j. ( Ed.), Sanchez, N. ( Ed.): Field Theory, Quantum Gravity
  and Strings*, 107-126 (1986).
  %\href{http://www.slac.stanford.edu/spires/find/hep/www?irn=1654543}{SPIRES entry}

\bibitem{BR}
V.~F.~Mukhanov, L.~R.~W.~Abramo and R.~H.~Brandenberger,
  ``On the back reaction problem for gravitational perturbations,''
  Phys.\ Rev.\ Lett.\  {\bf 78}, 1624 (1997)
  [arXiv:gr-qc/9609026];\\
  %%CITATION = GR-QC 9609026;%%
L.~R.~W.~Abramo, R.~H.~Brandenberger and V.~F.~Mukhanov,
  ``The energy-momentum tensor for cosmological perturbations,''
  Phys.\ Rev.\ D {\bf 56}, 3248 (1997)
  [arXiv:gr-qc/9704037].
  %%CITATION = GR-QC 9704037;%%

\bibitem{GB2}
B.~R.~Greene, T.~Prokopec and T.~G.~Roos,
  ``Inflaton decay and heavy particle production with negative coupling,''
  Phys.\ Rev.\ D {\bf 56}, 6484 (1997)
  [arXiv:hep-ph/9705357];\\
  %%CITATION = HEP-PH 9705357;%%
J.~Garcia-Bellido and A.~D.~Linde,
  ``Preheating in hybrid inflation,''
  Phys.\ Rev.\ D {\bf 57}, 6075 (1998)
  [arXiv:hep-ph/9711360].
  %%CITATION = HEP-PH 9711360;%%

\bibitem{tachyonic2}
G.~N.~Felder, J.~Garcia-Bellido, P.~B.~Greene, L.~Kofman, A.~D.~Linde and I.~Tkachev,  ``Dynamics of symmetry breaking and tachyonic preheating,''
  Phys.\ Rev.\ Lett.\  {\bf 87}, 011601 (2001)
  [arXiv:hep-ph/0012142].
  %%CITATION = HEP-PH 0012142;%%

\bibitem{DK}
A.~D.~Dolgov and D.~P.~Kirilova,
  ``Production Of Particles By A Variable Scalar Field,''
  Sov.\ J.\ Nucl.\ Phys.\  {\bf 51}, 172 (1990)
  [Yad.\ Fiz.\  {\bf 51}, 273 (1990)].
  %%CITATION = SJNCA,51,172;%%

\bibitem{COBE}
G.~F.~Smoot {\it et al.},
  ``Structure in the COBE DMR first year maps,''
  Astrophys.\ J.\  {\bf 396}, L1 (1992).
  %%CITATION = ASJOA,396,L1;%%

\bibitem{Natalia}
N.~Shuhmaher and R.~Brandenberger,
  ``Non-perturbative instabilities as a solution of the cosmological moduli
  problem,''
  Phys.\ Rev.\ D {\bf 73}, 043519 (2006)
  [arXiv:hep-th/0507103].
  %%CITATION = HEP-TH 0507103;%%

\bibitem{BKS}
M.~Bastero-Gil, S.~F.~King and J.~Sanderson,
  ``Preheating in supersymmetric hybrid inflation,''
  Phys.\ Rev.\ D {\bf 60}, 103517 (1999)
  [arXiv:hep-ph/9904315].
  %%CITATION = HEP-PH 9904315;%%

\bibitem{DL}
A.~D.~Dolgov and A.~D.~Linde,
  ``Baryon Asymmetry In Inflationary Universe,''
  Phys.\ Lett.\ B {\bf 116}, 329 (1982).
  %%CITATION = PHLTA,B116,329;%%

\bibitem{AFW}
L.~F.~Abbott, E.~Farhi and M.~B.~Wise,
  ``Particle Production In The New Inflationary Cosmology,''
  Phys.\ Lett.\ B {\bf 117}, 29 (1982).
  %%CITATION = PHLTA,B117,29;%%

\bibitem{LL} L. Landau and Lifshitz, {\it Mechanics} (Pergamon, Oxford, 1960).

\bibitem{Arnold} V. Arnold, {\it Mathematical Methods of Classical Mechanics}
   (Springer, New York, 1978).

\bibitem{BSV}
  M.~Bershadsky, C.~Vafa and V.~Sadov,
  ``D-Strings on D-Manifolds,''
  Nucl.\ Phys.\ B {\bf 463}, 398 (1996)
  [arXiv:hep-th/9510225].
  %%CITATION = HEP-TH 9510225;%%

\bibitem{Sb}
  A.~Strominger,
  ``Massless black holes and conifolds in string theory,''
  Nucl.\ Phys.\ B {\bf 451}, 96 (1995)
  [arXiv:hep-th/9504090].
  %%CITATION = HEP-TH 9504090;%%

\bibitem{Minos}
M.~Axenides, R.~H.~Brandenberger and M.~S.~Turner,
  ``Development Of Axion Perturbations In An Axion Dominated Universe,''
  Phys.\ Lett.\  B {\bf 126}, 178 (1983).
  %%CITATION = PHLTA,B126,178;%%

\bibitem{Gordon}
C.~Gordon, D.~Wands, B.~A.~Bassett and R.~Maartens,
  ``Adiabatic and entropy perturbations from inflation,''
  Phys.\ Rev.\  D {\bf 63}, 023506 (2001)
  [arXiv:astro-ph/0009131].
  %%CITATION = PHRVA,D63,023506;%%

\bibitem{BBDD} P. Brax, C. van de Bruck, A.-C. Davis and S. Davis,
   ``Coupling Hybrid Inflation to Moduli'',
   hep-th/0606140.
   %%CITATION = HEP-TH 0606140;%%


\end{thebibliography}
\end{document}